\documentclass[12pt]{article}
\pdfoutput=1
\usepackage[a4paper]{geometry}
\usepackage{jheppub, amsmath,amssymb,amsfonts,amsxtra, mathrsfs, makeidx,graphics,graphicx,amsthm,epsfig, ytableau,bm,longtable,float, color,tikz,mathtools,xfrac,footnote,rotating, lscape, makecell, environ,mathtools, empheq, multirow}

\usetikzlibrary{positioning}
\usetikzlibrary{chains}
\usetikzlibrary{arrows,fit,decorations.pathreplacing}
\tikzstyle{every picture}+=[remember picture]
\tikzstyle{na} = [baseline]

\usetikzlibrary{arrows, decorations.markings, calc, fadings, decorations.pathreplacing, patterns, decorations.pathmorphing, positioning}

\def\node#1#2{\overset{#1}{\underset{#2}{{\color{gray} \bullet}}}}

\def\sqbnode#1#2{\overset{#1}{\underset{#2}{{\color{blue} \blacksquare}}}}

\def\cver#1#2{\overset{{\llap{$\scriptstyle#1$}\displaystyle{\color{gray} \bullet}{\rlap{$\scriptstyle#2$}}}}{\scriptstyle\vert}}

\def\uer#1#2{\underset{{\llap{$\scriptstyle#1$}\displaystyle{\color{gray} \bullet}{\rlap{$\scriptstyle#2$}}}}{\scriptstyle\vert}}

\usepackage{bbm}
\usepackage{subfigure}

\newcommand{\eg}{\textit{e.g.}}

\newcommand{\ie}{\textit{i.e.}}

\numberwithin{equation}{section}

\newcommand{\nn}{\nonumber}

\newcommand{\be}{\begin{equation}} \newcommand{\ee}{\end{equation}}
\newcommand{\bea}{\begin{equation} \begin{aligned}} \newcommand{\eea}{\end{aligned} \end{equation}}

\def\tilde{\widetilde}

\def\hat{\widehat}

\def\rt2{\sqrt{2}}

\def\tr{\mathop{\rm tr}}

\def\CN{{\cal N}}

\def\CS{{\cal S}}
\def\CT{{\cal T}}

\def\1{{\ds 1}}

\newcommand{\bC}{\mathbb{C}}

\newcommand{\bR}{\mathbb{R}}
\newcommand{\bZ}{\mathbb{Z}}

\def\repa{\raise4pt\hbox{$\square$}\mkern-14mu\raise-4pt\hbox{$\square$}}
\def\repab{\overline{\raise4pt\hbox{$\square$}\mkern-14mu\raise-4pt\hbox{$\square$}\mkern-1mu}}

\def\smileface{\ensuremath{\hbox{\large$\bigcirc$}\mkern-15mu\raise-1pt\hbox{\scriptsize$\smallsmile$}%
\mkern-10mu\raise4pt\hbox{..}\mkern4mu}}
\def\frownface{\ensuremath{\hbox{\large$\bigcirc$}\mkern-15mu\raise-1pt\hbox{\scriptsize$\smallfrown$}%
\mkern-10mu\raise4pt\hbox{..}\mkern4mu}}

\newcommand{\ba}{\begin{array}}
\newcommand{\ea}{\end{array}}
\newcommand{\bi}{\begin{itemize}}
\newcommand{\ei}{\end{itemize}}

\def\bea#1\eea{\allowdisplaybreaks \begin{align}#1\end{align}}
 \newcommand{\ben}{\begin{enumerate}}
\newcommand{\een}{\end{enumerate}}
\newcommand{\bean}{\begin{eqnarray*}}
\newcommand{\eean}{\end{eqnarray*}}
\newcommand{\eref}[1]{(\ref{#1})}

\newcommand{\BC}{\mathbb{C}}

\newcommand{\BZ}{\mathbb{Z}}

\newcommand{\BH}{\mathbb{H}}

\newcommand{\comment}[1]{}
\newcommand{\diag}{\mathrm{diag}}

\newcommand{\blue}{\color{blue}}

\newcommand{\red}{\color{red}}

\newcommand{\bdash}{{\blue\rule[0.5ex]{.6cm}{0.6pt}}}
\newcommand{\rdash}{{\red\rule[0.5ex]{.6cm}{0.6pt}}}

\def\node#1#2{\overset{#1}{\underset{#2}{\circ}}}

\def\rnode#1#2{\overset{#1}{\underset{#2}{{\color{red} \bullet}}}}
\def\bnode#1#2{\overset{#1}{\underset{#2}{{\color{blue} \bullet}}}}

\def\sqbnode#1#2{\overset{#1}{\underset{#2}{{\color{blue} \blacksquare}}}}

\def\rver#1#2{\overset{{\llap{$\scriptstyle#1$}\displaystyle{\color{red} \bullet}{\rlap{$\scriptstyle#2$}}}}{\scriptstyle\vert}}

\def\bluesqver#1#2{\overset{{\llap{$\scriptstyle#1$}\displaystyle{\color{blue} \blacksquare}{\rlap{$\scriptstyle#2$}}}}{\scriptstyle\vert}}

\tikzstyle{every picture}+=[remember picture]
\tikzstyle{na} = [baseline=-.5ex]

\title{Small instanton transitions for M5 fractions}
\author[a,b]{Noppadol Mekareeya,}
\author[c,d]{Kantaro Ohmori,}
\author[e]{Hiroyuki Shimizu,}
\author[a,b]{and Alessandro Tomasiello}

\affiliation[a]{Dipartimento di Fisica, Universit\`a di Milano-Bicocca, \\ Piazza della Scienza 3, I-20126 Milano, Italy}
\affiliation[b]{INFN, sezione di Milano-Bicocca, I-20126 Milano, Italy}
\affiliation[c]{Department of Physics, The University of Tokyo, \\
Hongo, Bunkyo-ku, Tokyo 133-0022, Japan}
\affiliation[d]{School of Natural Sciences, Institute for Advanced Study, \\
1 Einstein Dr. Princeton, NJ 08540, USA}
\affiliation[e]{Kavli Institute for the Physics and Mathematics of the Universe, \\ University of Tokyo, Kashiwa, Chiba 277-8583, Japan}

\abstract{M5-branes on an ADE singularity are described by certain six-dimensional ``conformal matter'' superconformal field theories. Their Higgs moduli spaces contain information about various dynamical processes for the M5s; however, they are not directly accessible due to the lack of a Lagrangian formulation. Using anomaly matching, we compute their dimensions. The result implies that M5 fractions can recombine in several different ways, where the M5s are leaving behind frozen versions of the singularity. The anomaly polynomial gives hints about the nature of the freezing. We also check the Higgs dimension formula by comparing it with various existing conjectures for the CFTs one obtains by torus compactifications down to four and three dimensions. Aided by our results, we also extend those conjectures to compactifications of theories not previously considered. These involve class $\CS$ theories with twisted punctures in four dimensions, and affine-Dynkin-shaped quivers in three dimensions.}

\begin{document}
\maketitle

\section{Introduction} 
\label{sec:intro}
The dynamics of M5-branes is one of the most mysterious corners of string theory. It is described by a six-dimensional ${\cal N}=(2,0)$ superconformal field theory (SCFT) without a known Lagrangian description. 

A possible way to attack the problem is to see how the M5s interact with other objects. In particular, we can study how a stack of $N$ M5s behaves on a $\bR^4/\Gamma_G\times \bR$ singularity, where $\Gamma_G$ is a discrete subgroup of $SU(2)$ (associated to a Lie group $G$ by the McKay correspondence). Efforts in this direction are very old, but have been revived more recently thanks to progress in F-theory and holography. The resulting SCFT, which is denoted by ${\cal T}_{G}(N-1)$, has ${\cal N}=(1,0)$ supersymmetry and a $G\times G$ flavor symmetry. For a $\Gamma_G=\bZ_k$ singularity \cite{intriligator-6d,Intriligator:1997dh} one can obtain an effective description, valid on the tensor branch, the locus where the scalars $\phi_i$ in the tensor multiplets have generic vevs. Those tensors are coupled to a chain of $G=SU(k)$ gauge groups and bifundamental hypermultiplets. For $\Gamma_G=D_k$, $G=SO(2k)$, one gets an alternating sequence of $USp(2k-8)$ and $SO(2k)$ gauge groups. For $G=E_k$, one can obtain an effective description by using a dual F-theory description involving seven-branes wrapping a chain of curves. It consists of a more exotic sequence of gauge groups (called ``conformal matter'' in \cite{DelZotto:2014hpa}), also coupled to copies of the so-called E-string theory, a theory with one tensor multiplet and $E_8$ flavor symmetry. Both in the $D_k$ and in $E_k$ examples, the number of tensor multiplets is a multiple of the number of M5-branes; this is naturally interpreted as the fact that each M5 can break in several fractional M5s \cite{DelZotto:2014hpa}. (In the $D_k$ case, this was observed earlier in the IIA frame using NS5s on O6-planes \cite{evans-johnson-shapere}.) 

In the IIA duality frame, it is also possible \cite{Brunner:1997gf,hanany-zaffaroni-6d} to modify the theories ${\cal T}_{SU(k)}(N)$ by introducing D8s; these additional theories can be interpreted \cite{Gaiotto:2014lca} as the result of giving nilpotent vevs to some fields, thus activating a Higgs mechanism which breaks the $SU(k)\times SU(k)$ flavor symmetry. The resulting theories, which are denoted by ${\cal T}_{SU(k)}(\{Y_{\rm L}, Y_{\rm R}\},N)$, are labeled by two nilpotent elements, one for each copy of $SU(k)$ in the flavor symmetry group. The AdS$_7$ duals to these theories were found in \cite{Apruzzi:2013yva,10letter,Cremonesi:2015bld}. There is also an F-theory dual realization, where the nilpotent vevs appear as the residues for poles of the Hitchin equation living on some seven-branes \cite{DelZotto:2014hpa}, making them examples of ``T-branes'' \cite{Cecotti:2010bp}. The F-theory realization is more general: it allows us to find these T-brane theories ${\cal T}_G(\{Y_{\rm L}, Y_{\rm R}\},N)$ where the IIA picture breaks down, which includes many $D_k$ cases and all $E_k$ ones \cite{Heckman:2016ssk}. In the latter case the nilpotent elements $Y_{\rm L,R}$ are classified by so-called ``Bala--Carter labels'' (see \eg~ \cite{Chacaltana:2012zy}). 

The dimension $d_H$ of the Higgs moduli space for all these T-brane theories was studied in \cite{Mekareeya:2016yal}, both as a check of the F-theoretic predictions and as a way to better understand  the physics described by these theories. Using anomaly considerations, it was possible to show that the difference in $d_H$ between two T-brane theories ${\cal T}_G(\{Y_{\rm L}, Y_{\rm R}\},N)$, ${\cal T}_G(\{Y_{\rm L}',Y_{\rm R}'\},N)$ was equal to the difference $d_{Y_{\rm L}}+d_{Y_{\rm R}}-d_{Y_{\rm L}'}-d_{Y_{\rm R}'}$, where $d_Y$ is the dimension of the nilpotent orbit of $Y$ (namely, of the space of all nilpotent elements in $G$ conjugated to $Y$). This provided a strong cross-check of the F-theory predictions for T-brane theories. In \cite{Mekareeya:2016yal}, however, only differences of $d_H$ were computed, and not $d_H$ itself. This was because the Higgs moduli space one computes from an effective description on the tensor branch, $d_H^{\rm eff}$, is not always the same as the Higgs moduli space $d_H^{\rm CFT}$ of the CFT, which lives at the origin of the tensor branch, where the tensor multiplets scalars $\phi_i=0$. In other words, in general there can be a dimension jump: the Higgs moduli space lives at every point of the tensor branch, and its dimension has a generic value $d_H^{\rm eff}$ on a generic tensor branch point, while it can have a larger dimension on non-generic points, and an even higher value $d_H^{\rm CFT}$ at the CFT point at the origin. (To give justice to this stratified structure, the expression ``Higgs branch'' is a bit misleading. We will however use it in this paper for convenience.) This dimension jump issue is interesting as a field theory issue, but also because it has to do with what happens when M5 fractions meet each other --- or in other words with strongly coupled M5 dynamics.

In this paper we address this problem, and compute $d_H^{\rm CFT}$. We start by considering the original theories ${\cal T}_{G}(N)$. It is possible to pull the M5s away from the $\Gamma_G$ singularity: in other words, there should exist a flow from ${\cal T}_G(N)$ to a ${\cal N}=(2,0)$ theory, plus $d_H^{\rm CFT}$ additional free hypermultiplets. This property was called ``Higgsable to ${\cal N}=(2,0)$'' (HN) in \cite{Ohmori:2015pua}; here we use again anomaly matching to find that it imposes constraints on the field content, and thus on $d_H^{\rm CFT}$. The result reads
\begin{equation}\label{eq:dCFTintro}
	d_H^{\rm CFT}= n_H-n_V+29 n_{\rm GS} \, ,\qquad n_{\rm GS}\equiv\sum_{i,j} \eta^{-1}_{ij} (2- \eta^{ii})(2- \eta^{jj})\, .
\end{equation}
Here $n_H$ and $n_V$ are the numbers of hypermultiplets and vector multiplets respectively; the contribution $n_H-n_V$ is thus what one expects from field theories in other dimensions. $n_{\rm GS}$ is a Green--Schwarz-like contribution to a certain gravitational anomaly, in terms of $\eta$, the adjacency matrix of the theory, to be reviewed below. (In the F-theory realization, $\eta$ is simply the intersection matrix of the seven-branes; in purely field-theoretical terms, this summarizes the string charges of an instanton string of each individual gauge group, and how they are coupled.) 
For ${\cal T}_G(\{Y_{\rm L}, Y_{\rm R}\},N)$, (\ref{eq:dCFTintro}) gives
\begin{equation}\label{eq:dCFTTintro}
	d_H^{\rm CFT} = N+1+ {\rm dim}(G) - d_{Y_{\rm L}}- d_{Y_{\rm R}}\, .
\end{equation}

While ${\cal T}_G(N)$ satisfy the HN constraints by construction, there are other interesting examples. These consist of theories where the M5 fractions get reassembled in several different ways, leaving behind a ``frozen'' \cite{witten-without,deBoer:2001px,Ohmori:2015pua,tachikawa-frozen} (or partially so) version of the $\Gamma_G$ singularities, where the flavor gauge group becomes a group $G_{\rm fr}$ which can now be non-simply-laced (or empty). In such processes, some tensor multiplets are lost, and hypermultiplets are generated. (They are sometimes called ``small instanton transitions'', since they were first observed to happen for M5-branes turning into small instantons on an $E_8$ wall.)  We denote the SCFT associated with $N$ M5-branes on the (partially) frozen singularity $G_{\rm fr}$ by ${\cal T}^{\rm fr}_{G\to G_{\rm fr}}(N-1)$.  In fact, by using the classification in \cite{Heckman:2013pva}, we find that all examples of the HN theories are of this type (possibly again decorated by nilpotent elements). For these theories ${\cal T}^{\rm fr}_{G\to G_{\rm fr}}(N)$, the result (\ref{eq:dCFTTintro}) is still valid, with ${\rm dim}(G)$ replaced by ${\rm dim}(G_{\rm fr}).$

As a cross-check, we also compare our results with the Higgs moduli space dimensions of the theories one expects upon torus dimensional reduction. Indeed, explicit proposals were made \cite{Ohmori:2015pua,Ohmori:2015pia,Ohmorithesis2015} for the $T^2$ compactifications of 
the $G=SU(k)$ and $G=SO(2k)$ T-brane theories, in terms of class $\CS$ theories. For example, the particular case ${\cal T}_{G}(0)$, describing a single M5 on a $\Gamma_G$ singularity, was identified in \cite{Ohmori:2015pua} as a class $\CS$ theory of type $G$ associated to a sphere with one minimal and two maximal punctures. Our Higgs dimension matches with what one computes from these various four-dimensional proposals, as one expects. 

Drawing lessons from this computation, we can also propose natural generalizations of \cite{Ohmori:2015pua,Ohmori:2015pia,Ohmorithesis2015} to ${\cal T}_{E_k}(\{Y_{\rm L}, Y_{\rm R}\},N)$, or to ${\cal T}^{\rm fr}_{G\to G_{\rm fr}}(N)$, for non-simply-laced $G_{\rm fr}$. Both still involve class $\CS$ theories. In the non-simply-laced case, we make use of theories with twisted punctures, arising from a fixture of a larger simply-laced group $G$. While these proposals are a little less explicit than the cases in  \cite{Ohmori:2015pua,Ohmori:2015pia,Ohmorithesis2015}, they still pass the test of Higgs dimension comparison. 

A particularly interesting case is when $G_{\rm fr}$ is trivial, $G_{\rm fr}=\{1\}$ (which by an abuse of language we will denote by $\varnothing$); namely, the completely frozen case. For these theories, (\ref{eq:dCFTTintro}) gives the CFT Higgs branch dimension of ${\cal T}^{\rm fr}_{G\to \varnothing}(N-1)$ to be $N$. The $T^3$ compactification of ${\cal T}^{\rm fr}_{G\to \varnothing}(0)$ (\ie~one M5-brane) also has a Coulomb branch equal to the dual Coxeter number of $G$. This suggests the Higgs moduli space is simply a symmetric product of $\bC^2/\Gamma_G$, and gives in turn a natural conjecture for the $T^3$ compactification in terms of quivers with the shape of the extended Dynkin diagram of $G$. Both this and the $T^2$ compactifications discussed earlier give examples of applications of our formulas (\ref{eq:dCFTintro}), (\ref{eq:dCFTTintro}).
 
In section \ref{sec:fm5} we will give a short review of how six-dimensional theories are realized in F-theory; we will also review how it leads to the theory of fractional M5s, and define the partially frozen versions of these theories.
In section \ref{sec:classes} we will define some of the terminology we will use in the rest of the paper. In section \ref{sec:cons} we derive the constraints that follow from those definitions and anomaly matching, and use them to derive (\ref{eq:dCFTintro}). We then apply this result in section \ref{sec:M5} to point out some simple results about M5 recombination. In sections \ref{sec:sl-cm}, \ref{sec:T} and \ref{sec:nonsimplylacedcm} we look at $T^2$ compactifications of our theories, and in section \ref{sec:frozen} at $T^3$ compactifications of completely frozen theories.


\section{F-theory and fractional M5-branes} 
\label{sec:fm5}

We will start by reviewing some aspects of how F-theory engineers six-dimensional theories and fractional M5-branes. In the last subsection we will introduce frozen conformal matter, and characterize it. 

\subsection{6d theories from F-theory} 
\label{sub:f}

We will use at various points the language of F-theory to describe our six-dimensional theories. Let us quickly review the basics of this; for more information see for example \cite{Heckman:2013pva,Heckman:2015bfa}. Each theory is engineered by a tree (in fact, most often a chain) of seven-branes wrapping non-trivial curves. Each curve is characterized by its self-intersection, and by the gauge algebra, which is dictated by the Kodaira degeneration of the elliptic fibration over the curve. This is usually summarized by a diagram where a curve is represented by an integer $1\le n\le 12$ (representing self-intersection $-n$) with the gauge algebra on top. For example $\overset{\mathfrak{su_{2}}}2$ represents a curve with self-intersection number $-2$ and with gauge algebra $\mathfrak{su_2}$. There can also be ``empty'' curves without a gauge algebra: this can happen for $n=1$ or $2$. For $n=2$, this represents the $\CN=(2,0)$ theory of type $A_1$; for $n=1$, it represents the so-called ``E-string theory'', which has a tensor but also an $E_8$ flavor symmetry. Finally, non-compact curves may also be present; their would-be gauge algebra is then a flavor symmetry, and is usually denoted within square brackets.   
An elaborate example where all these features are present is   
\begin{equation}\label{eq:2E8cm}
	[E_8]\,\,1\,\,2 \,\, \overset{\mathfrak{su_{2}}}2 \,\, \overset{\mathfrak{g_{2}}}3  \,\, 1 \,\, \overset{\mathfrak{f_{4}}}5 \,\,  1 \,\, \overset{\mathfrak{g_{2}}}3 \,\, \overset{\mathfrak{su_{2}}}2 \,\, 2 \,\, 1\,\,  \overset{\mathfrak{e_{8}}}{12}  \,\,  1\,\, 2 \,\, \overset{\mathfrak{su_{2}}}2 \,\, \overset{\mathfrak{g_{2}}}3  \,\, 1 \,\, \overset{\mathfrak{f_{4}}}5 \,\,  1 \,\, \overset{\mathfrak{g_{2}}}3 \,\, \overset{\mathfrak{su_{2}}}2 \,\, 2 \,\, 1 \, \,[E_8]
\end{equation}
which is actually an example of the conformal matter theories mentioned in the introduction: it describes 2 M5s on a $\Gamma_{E_8}$ singularity, each of which has broken down in 12 fractions. There is an $E_8\times E_8$ flavor symmetry, each carried by an E-string theory, represented by an empty $(-1)$-curve. There are also several more copies of the E-string in (\ref{eq:2E8cm}). The two $(-1)$-curves touching the $\overset{\mathfrak{e_{8}}}{12}$ curve are E-string theories whose $E_8$ flavor symmetry has been fully gauged; the other four $(-1)$-curves touch each a $\mathfrak{g_2}$ and a $\mathfrak{f_4}$ gauge algebra, which means that a subalgebra $\mathfrak{g_2}\oplus\mathfrak{f_4}$ of the $E_8$ flavor has been gauged.

So far we have described a 6d theory on a generic point of its tensor branch. It is sometimes useful to consider a non-generic locus that one can obtain by ``blowing down'' the $-1$ curves, that is by shrinking them to zero size. A classic algebraic-geometrical result says that the self-intersection of the neighboring curves changes by 1: namely, $\ldots n1m\ldots \to \ldots(n-1)(m-1)\ldots$. This might create new $(-1)$-curves, that can be shrunk as well. The locus on the tensor branch of the theory where there are no longer any $(-1)$-curve is called the ``endpoint''. For example for (\ref{eq:2E8cm}) it can be checked that the endpoint consists of a single compact curve:
\begin{equation}
	[E_8]\,\,\overset{\mathfrak{e_{8}}}{2}\,\,[E_8]\ .
\end{equation}
It represents the locus where the 24 M5 fractions have coalesced in two full M5.  

In \cite{Heckman:2013pva}, a classification of all possible endpoints for a 6d SCFT was obtained. This lumped many theories together, and a finer classification of 6d SCFTs was then presented in \cite{Heckman:2015bfa}. 

Finally, we will need the concept of adjacency matrix $\eta$ associated to an F-theory chain. This is an $n_T\times n_T$ matrix, where $n_T$ is the number of curves in the chain, defined by 
\begin{equation}
	\eta^{ij}= \left\{ \begin{array}{cc}
		n_i &\qquad  i=j~,\\
		-1 &\qquad  |i-j|=1\, ,
	\end{array}\right.
\end{equation}
where $n_i$ is minus the self-intersection number for the $i$-th curve in the chain. One can apply this definition also to the endpoint; accordingly, we will call $\eta^{\rm end}$ the resulting $n_T^{\rm end}\times n_T^{\rm end}$-matrix.


\subsection{M5 fractionation} 
\label{sub:m5}

A stack of $N$ M5-branes at a $\bR\times \bR^4/\Gamma_G$ singularity is described by a ${\cal N}=(1,0)$ CFT ${\cal T}_G(N-1)$. Via a sequence of dualities, this can be described in terms of F-theory \cite{DelZotto:2014hpa}. Naively one would expect $N$ tensor multiplets, whose scalars would parameterize  the $N$ positions along the $\bR$ direction. At a generic point of this tensor branch, one would then expect each segment of singularity to give rise to a $G$ vector multiplet, and bifundamental hypermultiplets connecting neighboring gauge groups. However, this picture cannot be quite correct: one reason is for $G\neq SU(k)$ that there is no possible definition of bifundamental representation allowed by anomaly cancellation. In the F-theory setup, the $G$ gauge groups are realized on seven-branes, and the problem manifests itself as the fact that a point of contact between two such branes has a singularity not present in Kodaira's classification. This can be cured by blowing up the contact point repeatedly, which result in a sequence of additional intermediate curves replacing it. The final result is
\be \label{eq:cm}
\begin{split}
G=SU(k): &\quad   [SU(k)]\overset{\mathfrak{su}_k}2 \,\,\ldots \,\, \overset{\mathfrak{su}_k}2  [SU(k)] \\
G=SO(2k): &\quad [SO(2k)] \, \, \overset{\mathfrak{usp}_{2k-8}}1 \,\, \overset{\mathfrak{so}_{2k}}4    \,\,\dots\,\, [SO(2k)] \\
G= E_6: &\quad [E_6] \,\, 1  \,\,   {\overset{\mathfrak{su_{3}}}3}  \,\,1\,\,   {\overset{\mathfrak{e_6}}6}  \,\, \ldots \,\, [E_6] \\
G= E_7: &\quad [E_7] \,\, 1  \,\,   {\overset{\mathfrak{su_{2}}}2}  \,\,{\overset{\mathfrak{so_{7}}}3}\,\, {\overset{\mathfrak{su_{2}}}2}\,\, 1\,\, {\overset{\mathfrak{e_7}}8}  \,\, \ldots \,\, [E_7] \\
G= E_8: &\quad [E_8] \,\, 1  \,\,2\,\,   {\overset{\mathfrak{su_{2}}}2}  \,\,{\overset{\mathfrak{g_2}}3}\,\,1\,\, {\overset{\mathfrak{f_4}}5}\,\, 1\,\, {\overset{\mathfrak{g_2}}3}
\,\,{\overset{\mathfrak{su_{2}}}2}  \,\, 2 \,\, 1 \,\,{\overset{\mathfrak{e_8}}12}\ldots \,\, [E_8]~,
\end{split}
\ee
with the understanding that each sequence of curves is repeated $N-1$ times (and in particular the total number of $G$ gauge factors is $N-1$). The two external copies of $[G]$ represent a $G\times G$ flavor symmetry. A condensed common notation that we will use sometimes is
\begin{equation}\label{eq:bdash}
	[G]\bdash (G)\bdash\ldots\bdash(G)\bdash[G]~.
\end{equation}
For example, the $N=2$, $G=E_8$ case is (\ref{eq:2E8cm}), which in the condensed notation we would write $[E_8]\bdash(E_8)\bdash[E_8]$. The endpoint of all these theories is a sequence of $N-1$ $(-2)$-curves. 

We see in (\ref{eq:cm}) that for $\mathfrak{g} \neq\mathfrak{su}_k$ two $\mathfrak{g}$ gauge algebras are connected by a chain with its own tensor branch, rather than by the tensor and bifundamental hypermultiplet that connect two $\mathfrak{su}_k$ gauge algebras. This chain is the field theory manifestation of the sequence of blowups we mentioned above; in \cite{DelZotto:2014hpa} it was called ``conformal matter theory''.\footnote{To be more specific, these chains are sometimes called $(G,G)$ conformal matter, to highlight that they can be used to connect two $G$ gauge groups. There also exist $(G,G')$ chains, some of which we will encounter in section \ref{sec:fracM5}. Nevertheless, we have chosen to denote the theories in (\ref{eq:cm}) by ${\cal T}_G(N)$, to obtain less heavy-looking formulas.} 
We see in particular that for  $\mathfrak{g} \neq\mathfrak{su}_k$ there is more than one tensor multiplet between two $\mathfrak{g}$ gauge algebras, rather than just one. 
As we mentioned, the point of view of the M-theory interpretation, the scalar in a tensor multiplet represents the motion of an M5 in the $\bR$. Since there are now several tensor multiplets, it is natural to conjecture that the M5 has now broken down in several fractions. For the $G= SO(2k)$ case, this has a IIA counterpart in the fact that NS5-branes can break in two fractions on an O6-plane \cite{evans-johnson-shapere}. For the $G=E_k$ cases, we see from (\ref{eq:cm}) that the number of fractions $\mathfrak{f}$ is 4 for $E_6$, 6 for $E_7$, 12 for $E_8$. Summarizing,
\begin{equation}\label{eq:fr}
	\mathfrak{f}(SU(k))=1 \, ,\qquad \mathfrak{f}(SO(2k))=2 \, ,\qquad \mathfrak{f}(E_6)= 4 \, ,\qquad \mathfrak{f}(E_7)=6 \, ,\qquad
	\mathfrak{f}(E_8)=12~. 
\end{equation}
 Another aspect of (\ref{eq:cm}) is that, upon crossing an M5 fraction, the gauge algebra $\mathfrak{g}$ gets broken to a subalgebra $\mathfrak{g}_{\rm fr}$ (possibly even trivial). Since this affects the possible resolutions of those singularities, this is called a ``frozen'' version of the singularity.

This ``freezing'' of the singularity is characterized by a discrete flux of the M-theory 3-form field $C$ \cite{deBoer:2001px,tachikawa-frozen}:
\begin{equation}
    \int_{S^3/\Gamma_G}C =\frac{n}{d} \qquad\text{ mod 1},
    \label{eq:Cflux}
\end{equation}
where $S^3/\Gamma_G$ denotes the orbifolded unit sphere around the singularity in $\mathbb{C}^2/\Gamma_G$, and $n,d$ are coprime. The denominator $d= d_{G \to G_{\rm fr}}$ is given by the following table (see \cite[Eq.~(1.1), Table 1]{tachikawa-frozen} and \cite[Table 14]{deBoer:2001px}):
\be \label{tab:d}
\begin{tabular}{|c|c|c|}
\hline
$G$ & $G_{\text{fr}}$ & $d_{G \rightarrow G_{\text{fr}}}$ \\
\hline
$SO(2k+8)$ & $USp(2k)$ & $2$ \\
\hline
\multirow{2}{*}{$E_6$} & $SU(3)$ & 2 \\
& $\varnothing$ & 3\\
\hline            
\multirow{3}{*}{$E_7$} & $SO(7)$ & 2 \\
            & $SU(2)$ & 3\\                   
			& $\varnothing$ & 4\\	
\hline          
\multirow{5}{*}{$E_8$}& $F_4$ & 2 \\
            & $G_2$ & 3\\
            & $SU(2)$ & 4\\    
            & $\varnothing$ & 5\\
			& $\varnothing$ & 6\\
\hline              
\end{tabular}
\ee
A fractional M5 brane is then a domain wall dividing the singular locus, and the value of the discrete flux \eref{eq:Cflux} can be different for each domain.
Let the singular locus be at $x^{8,9,10,11}=0$ and consider a fractional M5 which sits at $x^{7,8,9,10,11}=0$, between domain $x^7<0$ with discrete flux $r_1$ and domain $x^7>0$ with discrete flux $r_2\ge r_1$. Then, the M5 charge of the domain wall is
\begin{equation}
	\left|\int_{S}\mathrm{d}C\right|
	= \int_{\{\varepsilon\}\times S^3/\Gamma_G}C
	- \int_{\{-\varepsilon\}\times S^3/\Gamma_G}C
	=r_2-r_1,
	\label{eq:M5charge}
\end{equation}
where $\varepsilon>0$ and $S=[-\varepsilon,\varepsilon]\times S^3/\Gamma_G\cup \{\varepsilon\}\times D^3/\Gamma_G\cup \{-\varepsilon\}\times D^3/\Gamma_G$ is the 5-sphere surrounding the fractional M5.

In \cite{Ohmori:2015pua} the system was compactified on $T^3$ and dualized so that the M5s become M2s; the fractions were then shown to be transitions between ``gauge triples'' \cite{deBoer:2001px} around the $T^3$.


\subsection{T-brane theories} 
\label{sub:t}

Conformal matter theories can be decorated by adding on the two outermost curves a feature that does not modify the geometric F-theory data: a T-brane \cite{Cecotti:2010bp}. The transverse fluctuation of an F-theory seven-brane are parameterized by a Higgs field; making it nilpotent does not change the eigenvalues, and thus the position of the brane, but it does nevertheless have physical content. The type of T-brane relevant for SCFT$_6$s is a pole for this Higgs field. 

This possibility was originally suggested by duality with IIA configurations involving D8-branes \cite{DelZotto:2014hpa}. The $T_{SU(k)}(N)$ conformal matter theory has a realization in IIA as an NS5--D6 intersection. But in IIA one can also introduce D8s \cite{Brunner:1997gf,hanany-zaffaroni-6d}, and the combinatorics of how it is possible to introduce them while still getting a SCFT$_6$ are summarized by two Young diagrams $Y_{\rm L,R}$; let us then call these theories ${\cal T}_{SU(k)}(\{Y_{\rm L}, Y_{\rm R}\},N)$. This is particularly natural from the point of view of the Nahm equations living on the D6: the D8s represent poles for those equations, with a nilpotent residue; indeed nilpotent elements in $SU(k)$ are parameterized by Young diagrams. Going to IIB by a T-duality, this becomes a pole for the Higgs field on the seven-brane.\footnote{\label{foot:T} The nonabelian nature of this pole might suggest a Myers-like effect in the limit where the gauge algebras are large, but this is in fact naive \cite{Bena:2016oqr}.} This strongly suggests that in F-theory it should be possible to similarly decorate any conformal matter theory (even for $G\neq SU(k)$) by two nilpotent elements $Y_{\rm L,R}\in G$, obtaining a more general set of theories
\begin{equation}\label{eq:T}
	{\cal T}_G(\{Y_{\rm L}, Y_{\rm R}\},N)~.
\end{equation}

A tensor branch description of all the theories (\ref{eq:T}) was obtained in \cite{Heckman:2016ssk}, for any ADE Lie group $G$, by using a property they are expected to have, namely that they are connected to ${\cal T}_G(N)$ by a Higgs RG flow. It was shown there that the web of flows one obtains from ${\cal T}_G(N)$ is in bijective correspondence to the Hasse diagram of nilpotent elements. It was later shown in \cite{Mekareeya:2016yal} that the drop in moduli space dimension in such an RG flow is exactly equal to $d_{Y_{\rm L}}+d_{Y_{\rm R}}$, the sum of the dimensions of the nilpotent orbits associated to $Y_{\rm L,R}$. 

One can check from \cite{Heckman:2016ssk} that all the T-brane theories have all $2\ldots 2$ as an endpoint. This confirms their claimed origin as a decoration of the original sequence of curves by two poles with nilpotent residues, since nilpotent Higgs fields do not change the geometry (although see footnote \ref{foot:T}). It was argued in fact in \cite{Heckman:2016ssk} that all the theories with $2\ldots 2$ endpoint are T-brane theories, possibly up to short outliers.  

Unfortunately it is currently not clear what these T-brane decorations mean in the original M-theory duality frame. It would be very interesting to clarify this.


\subsection{Frozen conformal matter} 
\label{sub:fm5}

In this paper, we are also interested in ``incomplete'' versions of the chains in (\ref{eq:cm}) --- namely, to the chains that result from taking some of the outermost fractions (i.e.~tensor multiplets) to infinity. For example, for $G=E_7$ we can take to infinity the two outermost fractions on the left and the three outermost on the right, and we end up with  
\begin{equation}\label{eq:ex-inc}
	[SU(2)]  \,\,{\overset{\mathfrak{so_{7}}}3}\,\, {\overset{\mathfrak{su_{2}}}2}\,\, 1\,\, {\overset{\mathfrak{e_7}}8}  \,\, \ldots \,\, {\overset{\mathfrak{e_7}}8}\,\,1\,\,{\overset{\mathfrak{su_{2}}}2}\,\, [SO(7)] ~.
\end{equation}
The chain now ends on the left and on the right with a partially frozen version of the singularity, in the sense explained above. For this reason, we will sometimes call this general class of theories \textit{frozen conformal matter}.

This class is more general than the ordinary ``unfrozen'' conformal matter we reviewed earlier. The endpoint is no longer a sequence of $-2$ curves; for example, for (\ref{eq:ex-inc}) it is $232^{n-2}3$, where $n$ is the number of $\mathfrak{e_7}$ gauge algebras and $2^n\equiv \underbrace{2\ldots 2}_{n}$. 

If we look at the endpoints for all frozen conformal matter, we cover all the (non-bifurcated) endpoints which are a priori possible for any SCFT$_6$, as classified in \cite{Heckman:2013pva}. The general rule is easy to describe:
\begin{equation}\label{eq:end}
	a_1 \underbrace{(G)\bdash(G)\bdash\ldots (G)}_{\#\,\,\text{of}\,\,(G)\,\,=\,\,n}a_2^t \qquad \mapsto \qquad e(a_1) 2^{n-2}e(a_2)^t~,
\end{equation}
where ${}^t$ denotes inversion of order, and $e$ is described by table \ref{tab:end}. When $n=1$, (\ref{eq:end}) has to be understood according to the following rule: 
\begin{equation}\label{eq:2-1}
	\cdots x 2^{-1} y \cdots \equiv \cdots (x+y-2)\cdots~.
\end{equation}
The general rules (\ref{eq:end}) and (\ref{eq:2-1}) are enough to cover all the possible incomplete chains, except for those that do not include any copy of the ``original'' gauge algebra $\mathfrak{g}$ at all, which can be easily handled separately. 

\begin{table}[ht]
	\centering
$SO(2k):
$\begin{tabular}{|c|cc|}
\hline
$a$ & 1 &  $\varnothing$ \\
$e(a)$ & 2 &  3 \\
\hline
\end{tabular}	
\qquad $E_6:$
\begin{tabular}{|c|cccc|}
\hline
$a$ & 131 & 31 & 1 &  $\varnothing$ \\
$e(a)$ & 2 & 23 & 3 & 4\\
\hline
\end{tabular}\\\vspace{.2cm}
$ E_7:$
\begin{tabular}{|c|cccccc|}
\hline
$a$ & 12321 & 2321 & 321 & 21 & 1 & $\varnothing$ \\
$e(a)$ & 2 & 223 & 23 & 3 & 4 & 5\\
\hline
\end{tabular}
\\\vspace{.2cm}
$E_8:$
\begin{tabular}{|c|cccccc|}
\hline
$a$ & 12231513221 & 2231513221 & 231513221 & 31513221 & 1513221 & 513221 \\
$e(a)$ & 2 & 22223 & 2223 & 223 & 23 & 33 \\
\hline
$a$  & 13221 & 3221 & 221 & 21 & 1 &  $\varnothing$ \\
$e(a)$ & 3 & 24   & 4 & 5 & 6 & 7\\
\hline
\end{tabular}
\caption{The map $e$ used in the general endpoint result (\ref{eq:end}).}
\label{tab:end}
\end{table}

For example, the endpoint of (\ref{eq:ex-inc}) can be obtained using (\ref{eq:end}) as follows. In this case, $a_1= 321$, and $a_2^t=12$, or in other words $a_2=21$. Table \ref{tab:end} gives $e(a_1)=23$, $e(a_2)=3$, and then $e(a_2)^t=3$. So now the endpoint is $232^{n-2}3$, as previously stated. For $n=1$, we have to use (\ref{eq:2-1}), which gives $232^{-1}3= 24$. 
 
Looking at table \ref{tab:end}, we see that the $e(a)$ cover all the possible $\alpha$ in the endpoint classification \cite[Eq.~(5.7),(5.9)]{Heckman:2013pva};\footnote{$e(a)=$ 6, 5, 4, 3, 23 do not appear in their list of $\alpha$s because they can be obtained from 7 and 24: the list of $\alpha$s in \cite[Eq.~(5.9)]{Heckman:2013pva} provides the ``convex hull'' of endpoints, as explained there.} the $\beta$s there are simply $\alpha^t$ in our notation. Moreover, with the rule (\ref{eq:2-1}) we also cover all the ``rigid outliers'' of \cite[Eq.~(5.12)]{Heckman:2013pva}. 

Thus the frozen conformal matter we discussed in this subsection provide all the non-bifurcated endpoints in \cite{Heckman:2013pva} where the chain does not bifurcate. (In \cite{Heckman:2013pva} there are also a few endpoints which do bifurcate but quite minimally, listed in their (5.8), which we are not covering here. It is possible that one might obtain those by generalizing the present discussion by introducing orientifold-type objects.) We hasten to add that many different theories can share the same endpoint, so the frozen conformal matter theories are far from being the most general SCFT$_6$. 

However, as we mentioned at the end of section \ref{sub:t}, the theories with endpoint $2^n$ can be obtained from ordinary conformal matter by decorating them with nilpotent elements, thus obtaining the theories (\ref{eq:T}). This suggests that one could similarly obtain all the non-bifurcated theories by decorating frozen conformal matter by nilpotent elements. Some elements for doing so were already analyzed in \cite[Sec.~4.2]{Heckman:2016ssk}, where nilpotent hierarchies for $G_2$ and $F_4$ were obtained.



\section{Very Higgsable, and Higgsable to ${\cal N}=(2,0)$ theories}
\label{sec:classes}

We will now introduce some terminology, extending ideas introduced in \cite{Ohmori:2015pua, Ohmori:2015pia}.

\paragraph{Very Higgsable (vH) theories.} These are the CFTs whose Higgs branch is such that at its generic point the theory flows to a collection of free hypermultiplets, without any tensors:
\be \label{eq:tr}
\text{vH SCFT} \quad \rightarrow \quad \text{free hypermultiplets}~.
\ee

We divide this class in two subclasses: Obviously very Higgsable, which are strictly speaking the ones considered in \cite{Ohmori:2015pua}, and Hiddenly very Higgsable, which are a natural extension.

\ben
\item {\bf Obviously very Higgsable (OvH) theories.} These theories can be identified as very Higgsable
directly by looking at the F-theory realization: all of the compact cycles producing the tensor multiplets can be removed by repeated blow-downs of $(-1)$-curves. Examples of such theories include
\bi
\item The theory of free hypermultiplets.
\item The E-string theory we mentioned above, with one tensor and an $E_8$ flavor symmetry. It has a Higgs branch of dimension 29. One incarnation of this theory is as the description of an M5 on an $E_8$ boundary in M-theory; in this picture, the tensor branch corresponds to pulling away the M5 from the wall, while the Higgs branch corresponds to turning the M5 into an $E_8$ instanton. Since such an instanton cannot be pulled off the wall, we see that the tensor multiplet has been lost on this branch. This is sometimes called a ``small instanton transition''; by a slight abuse a language, one sometimes calls this way any transition where tensors are lost in favor of hypermultiplets, such as the flow in (\ref{eq:tr}). 
\item The so-called rank $N$ E-string theory, which consists of a single $(-1)$-curve followed by $N-1$ (-2)-curves, and which describes $N$ M5s on the $E_8$ wall. 
\item The worldvolume theory of a single or multiple M5-branes on $\BC^2/\Gamma_G$ on an $E_8$ wall.
\item The worldvolume theory ${\cal T}_G(0)$ of a single M5-brane on $\BC^2/\Gamma_G$, also known as the minimal conformal matter theory of type $(G,G)$, with $G$ a simply-laced group.
\item Certain theories that describe fractional M5-branes on orbifold singularities, including 
\bi
\item $(E_7, SO(7))$ minimal conformal matter describing $1/2$ M5-branes on $\BC^2/\Gamma_{E_7}$; 
\item $(E_8, F_4)$ minimal conformal matter describing $1/2$ M5-branes on $\BC^2/\Gamma_{E_8}$; and 
\item $(E_8, G_2)$ minimal conformal matter describing $1/3$ M5-branes on $\BC^2/\Gamma_{E_8}$.
\ei
\ei
\item {\bf Hiddenly very Higgsable (HvH) theories.} These are very Higgsable theories which are not obviously very Higgsable.  We will give an F-theory characterization in section \ref{sub:HvHF}. Examples of such theories include
\bi
\item  The worldvolume theory of a single M5-brane probing the completely frozen $SO(2k)$ or $E_{6,7,8}$ singularity.
\item  The theories describing a single M5-brane probing an ADE singularity frozen to a non-simply-laced group $G_{\rm fr}$; this will turn out to be $G_2$, $F_4$, $USp(2k)$. (Such theories are one possible definition of $(G_{\rm fr},G_{\rm fr})$ conformal matter.) We will see these in more detail in section \ref{sec:M5}.
\ei
\een

\paragraph{Higgsable to $\CN=(2,0)$ (HN) theories.} This is a generalization of the notion of vH theory. A theory is HN if its Higgs branch is such that at its generic point the theory flows to an ${\cal N}=(2,0)$ theory plus free hypermultiplets:
\begin{equation}\label{eq:HNflow}
\text{HN SCFT} ~ \rightarrow ~ \text{$\CN=(2,0)$ of type $\mathfrak{g}$}~ +~ \hat{d}\,\, \text{hypers}.
\end{equation}
We again divide the class in two subclasses: those for which such a property is obvious from the F-theory realization, which are strictly speaking the ones considered in \cite{Ohmori:2015pia}, and those for which such a property is hidden.

\ben
\item {\bf Obviously Higgsable to $\CN=(2,0)$ (OHN) theories.} In terms of F-theory, these theories have only $(-2)$-curves at the endpoint where all possible blow-downs of $(-1)$-curves have been performed. The number of $(-2)$-curves at the endpoint was denoted by $\mathfrak{n}$ in \cite{Mekareeya:2016yal} for this class of theories.  This number $\mathfrak{n}$ is precisely the rank of the $\CN=(2,0)$ theory at the endpoint of the aforementioned Higgs branch flow \cite{Ohmori:2015pia}.  Note that this class of theories were referred to as ``Higgsable to $\CN=(2,0)$'' in \cite{Ohmori:2015pua}. Examples are
\bi
\item The worldvolume theory of multiple M5-branes on $\BC^2/\Gamma_G$, also known as the non-minimal conformal matter of type $(G,G)$, with $G$ a simply-laced group. 
\item More generally, the T-brane theories that arise from the nilpotent Higgsing associated with a pair of the nilpotent orbits $(\mu_L, \mu_R)$ of a sufficiently long chain of the conformal matter of type $(G,G)$, with $G$ simply-laced. 
\ei 
\item {\bf Hiddenly Higgsable to $\CN=(2,0)$ (HHN) theories.}  After blowing-down all of the $(-1)$-curves, the endpoint does not consist of only $(-2)$-curves. The possible endpoints will be classified in section \ref{sub:HHNF}. We will also see that in a sense all such theories describe multiple M5-branes probing a (partially) frozen $SO(2k)$ or $E_{6,7,8}$ singularity (or their T-brane descendants). 
\een

We will consider  HvH and HHN theories in more detail in the next section.

\section{Constraints for the very Higgsable and Higgsable to $\CN=(2,0)$ theories}
\label{sec:cons}

In this section, we obtain a necessary condition for a 6d SCFT to be an obviously or hiddenly very Higgsable or Higgsable to $\CN=(2,0)$ theory.

\subsection{Very Higgsable theories}

During the flow (\ref{eq:tr}), the diffeomorphism group remains unbroken, and hence the gravitational anomalies can be matched in both sides.  

We compute the gravitational anomaly of the 6d SCFT by moving on to a sub-branch of the tensor branch where all $(-1)$-curves are blown-down, also known as the ``endpoint'' and classified by \cite{Heckman:2013pva}.  The tensor branch flow from the vH SCFT to the endpoint is as follows:
\be \label{vHflowtensor}
\text{vH SCFT}  \quad \rightarrow \quad \oplus_i \text{OvH}_i~ + ~\text{$n^{\text{end}}_V$ vectors} ~+~ \text{$n^{\text{end}}_H$ hypers}
\ee
where at the endpoint there are a collection of OvH theories (labelled $\text{OvH}_i$), $n^{\text{end}}_V$ vector multiplets and $n^{\text{end}}_H$ tensor multiplets.   The configuration of the tensor multiplets at the endpoint is specified by an integral, symmetric and positive definite matrix $\eta_{\text{end}}^{ij}$ with $i,j = 1, \ldots, n^{\text{end}}_T$, whose diagonal elements satisfy the inequality $2 \leq \eta_{\text{end}}^{ii} \leq 12$.  

The gravitational anomaly of 6d SCFT at the endpoint, computed using \eref{vHflowtensor}, is
\begin{equation}
I^{\text{end}} = I^{\text{\rm GS}} + n^{\text{end}}_T I^{\text{tensor}} +  n^{\text{end}}_V I^{\text{vector}} + \sum_i I^{\text{OvH}_i} \label{eq:grav},
\end{equation}
where the notation is as follows: 
\bi
\item The Green-Schwarz contribution $I^{\text{\rm GS}}$ to the gravitational anomaly at the endpoint is
\begin{equation}
\frac{n^{\text{end}}_{\rm GS}}{32} p_1(T)^2,
\end{equation}
where 
\begin{equation}\label{eq:ngs}
	n^{\text{end}}_{\rm GS} \equiv \sum_{i,j=1}^{n^{\text{end}}_T} (\eta^{-1}_{\text{end}})_{ij} (2- \eta^{ii}_{\text{end}})(2- \eta^{jj}_{\text{end}})\ . 
\end{equation}

\item The contribution of tensor/vector multiplet is given as 
\begin{align}
I^{\text{tensor}} = \frac{23p_1(T)^2 - 116 p_2(T)}{5760}, \qquad I^{\text{vector}} = - \frac{7p_1(T)^2 - 4 p_2(T)}{5760}.
\end{align}
\item The contribution $I^{\text{OvH}_i}$ is the gravitational anomalies of each $\text{OvH}_i$ theory on the right of \eref{vHflowtensor}.
\begin{equation}
I^{\text{OvH}_i} = d^{\text{OvH}_i}_H  \frac{7p_1(T)^2 -4p_2(T)}{5760}
\end{equation}
by using some positive integer $d^{\text{OvH}_i}_H$. This fact can be proven by the inductive method used in \cite{Ohmori:2015pua}. More precisely, the integer $d^{\text{end}}_H$ is written as
\begin{equation}
 d^{\text{OvH}_i}_H = 29 n^{\text{OvH}_i}_T +  n^{\text{OvH}_i}_H -  n^{\text{OvH}_i}_V
 \end{equation}
where $ n^{\text{OvH}_i}_{T,H,V}$ is the number of tensor/hyper/vector multiplets of the $\text{OvH}_i$ theory. 
\ei
Plugging these contributions back to \eqref{eq:grav}, we obtain the gravitational anomaly
\begin{equation} \label{grav1}
\begin{split}
&\frac{1}{5760}\left( 180n^{\text{end}}_{\rm GS}+23 n^{\text{end}}_T  - 7n^{\text{end}}_V + 7 \sum_i d^{\text{OvH}_i}_H \right)p_1(T)^2 \\ 
&- \frac{1}{5760} \left( 116n^{\text{end}}_T  - 4n^{\text{end}}_V  + 4 \sum_i d^{\text{OvH}_i}_H \right)p_2(T)~.
\end{split}
\end{equation}

Assume that on the right hand side of \eqref{eq:tr} there are $d_H$ hypermulitplets, where $d_H$ is the dimension of the Higgs branch of the vH SCFT in question at the origin of the tensor branch. The gravitational anomaly can also be written as 
\be \label{grav2}
I^{\text{end}} = d_H  \frac{7p_1(T)^2 -4p_2(T)}{5760}~.
\ee

Matching the gravitational anomalies \eref{grav1} and \eref{grav2}, we obtain two equations:
\begin{equation} \label{anommatch}
\begin{split}
180 n^{\text{end}}_{\rm GS}+23 n^{\text{end}}_T  - 7 n^{\text{end}}_V + 7 \sum_i d^{\text{OvH}_i}_H &= 7d_H, \\
116 n^{\text{end}}_T - 4 n^{\text{end}}_V + 4 \sum_i d^{\text{OvH}_i}_H &= 4d_H.
\end{split}
\end{equation}
From the second equation, we have 
\be \label{dHfromend}
d_H = 29 n^{\text{end}}_T  + \left( \sum_i d^{\text{OvH}_i}_H \right) - n^{\text{end}}_V ~. 
\ee
This equation has an interesting physical interpretation.  It implies that the Higgs branch dimension $d_H$ of the vH theory at the origin of the tensor branch can be computed using the endpoint data, including $n^{\text{end}}_T$ and $n^{\text{end}}_V$.  The quantity $\sum_i d^{\text{OvH}_i}_H$ should be viewed at the ``effective'' number of hypermultiplets at the endpoint.

By eliminating  $d_H$ from \eref{anommatch}, we obtain a nontrivial constraint for the possible tensor branch structure of the endpoint, namely
\begin{equation}
	n^{\text{end}}_T = n^{\text{end}}_{\rm GS} =  \sum_{i,j} (\eta^{-1}_{\text{end}})_{ij} (2- \eta^{ii}_{\text{end}})(2- \eta^{jj}_{\text{end}})\, , \label{eq:main}
\end{equation}
where we have recalled the definition of $n^{\text{end}}_{\rm GS}$ from (\ref{eq:ngs}). 
Note that if an the original vH theory is OvH, $n^{\text{end}}_T=0$ by definition; hence $ \sum_{i,j} (\eta^{-1}_{\text{end}})_{ij} (2- \eta^{ii}_{\text{end}})(2- \eta^{jj}_{\text{end}})=0$ for such a theory.

In fact, we can also obtain an equation similar to \eref{dHfromend}, but relating the Higgs branch dimension $d_H$ at the origin to data of the field theory at a generic point on the tensor branch.  Since the coefficient $p_2(T)$ is not affected by the Green-Schwarz contribution, this coefficient can be matched between the theory of $d_H$ free hypermultiplets and the effective field theory at a generic point on the tensor branch:
\be \label{dHfromgenerictensor}
d_H = 29 n_T +n_H -n_V
\ee
where $n_{T, H, V}$ are the number of the tensor multiplets, hypermultiplets and the vector multiplets at a generic point on the tensor branch.


\subsection{Characterization of HvH theories in terms of F-theory}
\label{sub:HvHF}

We will now characterize the solutions to the constraint (\ref{eq:main}) in F-theory, first by looking at some examples and then by giving the complete list. We will postpone the M-theory interpretation to section \ref{sec:M5}.

First of all, the constraint cannot be satisfied when the endpoint consists only of $-2$ curves, since the term on the right of the first equality of \eqref{eq:main} is zero. This prevents a number of 6d $\CN={(1,0)}$ theories from being very Higgsable.  For example, the worldvolume theories on multiple M5-branes on an orbifold singularity (non-minimal 6d conformal matter theories) does not belong to this class.

\paragraph{Theories with a single curve} Let us now turn to the theory of a single curve. If we denote denote the self-intersection of the curve as $-n$, the constraint \eqref{eq:main} becomes
\begin{equation}
\frac{1}{n}(n-2)^2 =1,
\end{equation}
whose solution is $n=1,4$. 

The solution $n=1$ corresponds to the rank-one E-string theory and is indeed OvH. As we reviewed earlier, the transition \eqref{eq:tr} results in 29 free hypermultiplets. 

The case of $n=4$ is more interesting. If we assume that the elliptic fibration over the $-4$ curve is as generic as possible, we obtain the gauge group $SO(8)$ and no hypermultiplet on the tensor branch; there is no Higgs branch at the generic point of the tensor branch. However, since $n^{\text{end}}_V=28$, $n^{\text{end}}_T=1$ and $d^{\text{OvH}_i}_H=0$, \eref{dHfromend} says that $d_H=1$, \ie~ there is a one dimensional Higgs branch at the origin of the tensor branch. Thus this theory is hiddenly very Higgsable. 

As we will see, there are several examples with this endpoint; two notable ones will appear in  sections \ref{sec:cmF4F4} and \ref{sec:frozen}. 
 
\paragraph{Theories with two curves} Let us consider a theory of two curves with the self-intersection $-n$ and $-m$. The intersection matrix and its inverse are given by
\begin{equation}
\eta = \begin{pmatrix}
m & -1 \\
-1 & n
\end{pmatrix}, \, \, \, \,  \eta^{-1} = \frac1{mn-1} \begin{pmatrix}
n & 1 \\
1 & m
\end{pmatrix}.
\end{equation}
Then, the constraint \eqref{eq:main} becomes 
\begin{equation}
\frac{mn (m+n-6)+8}{m n-1} = 2,
\end{equation}
whose solutions are $(m,n)=(1,2)$, $(1,5)$, $(2,5)$. The first two cases reduce to a single curve after blowing down the $(-1)$-curve; this was discussed in the previous paragraph. The solution $(2,5)$ is more interesting: in sections \ref{sec:cmG2G2} and \ref{sec:frozen}, we will discuss two examples of this type.

\paragraph{General analysis} After these preliminary examples, we can start a more systematic analysis. The constraint (\ref{eq:main}) quickly becomes complicated when one considers a number of curves $>2$. However, as mentioned in section \ref{sec:classes}, the full list of possible endpoints for a 6d SCFT was obtained in \cite{Heckman:2013pva}. So we can simply apply (\ref{eq:main}) to the list of endpoints in that paper.

The full list reads
\begin{equation}\label{eq:vHe}
	4 \, ,\qquad 52 \, ,\qquad 352 \, ,\qquad 622 \, ,\qquad 7222 \, ,\qquad 82222 \, . 
\end{equation}
We already made some comments about the first two cases. We will give a similar interpretation to the others in section \ref{sec:M5}.

\subsection{Higgsable to $\CN=(2,0)$ theories}
Recall that a theory is said to be Higgsable to $\CN=(2,0)$ (HN) if there is a Higgs flow (\ref{eq:HNflow}). For brevity, we write
\be
\mathfrak{n} = \mathrm{rank}\,\, \mathfrak{g}~.
\ee
Since the $\CN=(2,0)$ theory has a flow
\begin{equation}
 \text{$\CN=(2,0)$ of type $\mathfrak{g}$} ~ \to~ \text{$\mathfrak{n}$ copies of} ~(\text{$\CN=(1,0)$ hyper} + \text{$\CN=(1,0)$ tensor}),
\end{equation}
the HN SCFT in question has a flow 
\begin{equation}  \label{HNflow1}
\text{HN SCFT} ~\to~ d_H~\text{hypers} + \mathfrak{n}~\text{tensors}~,
\end{equation}
with 
\be
d_H=\hat{d}+\mathfrak{n}~.
\ee
We interpret $d_H$ as the dimension of the Higgs branch at the origin of the tensor branch.  We shall henceforth refer to $d_H$ as the CFT Higgs branch dimension.

Similarly to \eref{dHfromgenerictensor}, we can match the coefficient of $p_2(T)$ in the anomaly polynomials of the right hand side of \eref{HNflow1} with the effective field theory on the generic point of the tensor branch:
\begin{equation}\label{eq:dH}
	d_H + 29 \mathfrak{n} = 29 n_T+n_H -n_V~,
\end{equation}
where $n_{T, H, V}$ are the number of the tensor multiplets, hypermultiplets and the vector multiplets at a generic point on the tensor branch.  For $\mathfrak{n}=0$, we recover \eref{dHfromgenerictensor} for the vH theory as expected.

By matching the coefficient $p_1(T)^2$ of the anomaly polynomials of the right hand side of \eref{HNflow1} with the effective theory at the endpoint, we find that
\be \label{fraknnTend}
n^{\text{end}}_T = n^{\text{end}}_{\text{\rm GS}} + \mathfrak{n}  = \sum_{i,j} (\eta^{-1}_{\text{end}})_{ij} (2- \eta^{ii}_{\text{end}})(2- \eta^{jj}_{\text{end}}) + \mathfrak{n}\, ,
\ee
where again we have recalled the definition of $n^{\text{end}}_{\text{\rm GS}}$ from (\ref{eq:ngs}). This is a necessary condition for an SCFT to be HN: $n^{\rm end}_T-n^{\rm end}_{\rm GS}$ should be an integer.  For an OHN theory, the endpoint contains only $(-2)$-curves and the number of these curves are equal to $\mathrm{rank}(G)=\mathfrak{n}$ of the $\CN=(2,0)$ theory in question \cite{Ohmori:2015pia}; hence $ n^{\text{end}}_T=\mathfrak{n}$ and $ \sum_{i,j} (\eta^{-1}_{\text{end}})_{ij} (2- \eta^{ii}_{\text{end}})(2- \eta^{jj}_{\text{end}})=0$ for such a theory.

It also happens that $n_{\rm GS}- n_T$ is invariant under blowdown. So we have
\begin{equation}\label{eq:ngst-const}
	n_{\rm GS}^{\rm end}- n_T^{\rm end} = n_{\rm GS}- n_T~,
\end{equation}
where now $n_{\rm GS}=  \sum_{i,j=1}^{n_T} \eta^{-1}_{ij} (2- \eta^{ii})(2- \eta^{jj})$, which we already defined back in (\ref{eq:dCFTintro}). So the constraint (\ref{fraknnTend}) can also be imposed before flowing to the endpoint, by requiring that $n_{\rm GS}- n_T$ be an integer. 

In sections \ref{sec:nonsimplylacedcm} and \ref{sec:frozen}, we provide some examples of this class, including the non-minimal conformal matter theories of type $(G,G)$ where $G$ is non-simply-laced and theories on the worldvolume of multiple M5-branes on a completely frozen singularity. As a preview, here is an example of the latter: for two M5s on a frozen $D_4$ singularity, the quiver reads
\be
[1]\,\,\overset{\mathfrak{so}(8)}{4} \,\,  1 \,\,\overset{\mathfrak{so}(8)}{4}\,\,[1]
\ee
Indeed we can check that (\ref{fraknnTend}) is satisfied with $\mathfrak{n}=1$, which means that this theory has a Higgs branch flow to $\CN=(2,0)$ theory of type $\mathfrak{su}(2)$.

For an HN theory, there should be a tensor branch flow
\begin{equation}
\text{HN SCFT}~ \rightarrow~ \oplus_i  \mathrm{vH}_i + \mathfrak{n}\,\, \text{tensors} + \tilde{n}_V\,\, \text{vectors},
\end{equation}
where $\mathrm{vH}_i$ are collections of (obviously or hiddenly) vH theories.  Similarly to the above, the matching of the coefficient $p_2(T)$ yields
\begin{equation}
d_H=\sum_i d^{\mathrm{vH}_i}_H - \tilde{n}_V~.
\end{equation}

\subsection{Characterization of HHN theories in terms of F-theory}
\label{sub:HHNF}

Once again, to find the most general solution to (\ref{fraknnTend}), we can go through the list of endpoints provided by \cite{Heckman:2013pva} (just like we did for (\ref{eq:main}) at the end of section \ref{sub:HvHF}). 

The full list is given by 
\begin{equation}\label{eq:HNe}
	32^{\frak{n}-1}3 \, ,\qquad 42^{\frak{n}-1}32  \, ,\qquad 332^{\frak{n}-1}42  \, ,\qquad
	52^{\frak{n}-1}322 \, ,\qquad 62^{\frak{n}-1}3222 \, ,\qquad
	72^{\frak{n}-1}32222 \, ,
\end{equation}
where $2^{\frak{n}-1}\equiv \underbrace{2\ldots 2}_{\frak{n}-1}$, and $\mathfrak{n}$ coincides with what one computes from the constraint (\ref{fraknnTend}). Every $e(a)$ in table \ref{tab:end} appears on one side of an element of the list (\ref{eq:HNe}), with a companion on the other side (which is itself for $e(a)$=3).
Notice also that (\ref{eq:vHe}) can be obtained by (\ref{eq:HNe}) for $\frak{n}=0$ by using the rule (\ref{eq:2-1}). 

We will give an M5 interpretation to (\ref{eq:HNe}) in the next section.

\subsection{Summary} 
\label{sub:sum}

Let us summarize the main results from this section. An HN theory has to satisfy the constraint (\ref{fraknnTend}). Using (\ref{eq:dH}), (\ref{fraknnTend}) and (\ref{eq:ngst-const}), we can write the Higgs branch dimension of the CFT point as
\begin{equation}\label{eq:dCFT}
	\boxed{\dim^{\text{CFT}}_\BH ~\text{Higgs} = n_H-n_V + 29 n_{\rm GS} } 
\end{equation}
which is (\ref{eq:dCFTintro}), repeated here for convenience. $n_H$ and $n_V$ are the numbers of hypermultiplets and vector multiplets in the F-theory quiver at the generic point of the tensor branch.


In the following, we adopt the notation $\dim^{\text{CFT}}_\BH ~\text{Higgs}(\CT^{6d})$ to denote the Higgs branch dimension (in quarternionic units) of the CFT fixed point of the six dimensional theory $\CT^{6d}$.

\section{Fractional M5-branes}
\label{sec:M5}

We will now apply the results of the preceding section to M5-branes on singularities, and related theories.
 
\subsection{Conformal matter theories} 

As we already mentioned, these are OHN; $n_{\rm GS}=0$, and $\mathfrak{n}$ is equal to the number of $(-2)$-curves after blowing down all $(-1)$-curves. (\ref{eq:dCFT}) gives
\begin{equation}\label{eq:dCFTcm}
	\dim^{\text{CFT}}_\BH\text{Higgs}({\cal T}_G(\mathfrak{n}))= \mathfrak{n}+\dim(G)+1~.
\end{equation}

This formula has a physical interpretation. The theory is HN, and thus there is a transition where all the tensors disappear; however, they can disappear only on certain loci of the tensor branch. When some M5 fractions come together to form a full M5, the latter can be pulled away from the singularity. (In figure \ref{fig:M5-trans} we see a single M5 being formed in this fashion.) The positions of these full M5s parameterize the $\mathfrak{n}+1$ summand in (\ref{eq:dCFTcm}).

\subsection{T-branes} 

As we also already mentioned, all the T-brane theories  obtained in \cite{Heckman:2016ssk} are actually also HHN, although not especially obvious by their aspect at the generic point of the tensor branch. This is to be expected from their origin in F-theory: they are obtained by taking a conformal matter theory ${\cal T}_G(\mathfrak{n})$ and adding a pole for the Hitchin field at the two outermost -2 curves. This was already checked in \cite{Mekareeya:2016yal}. From the results in that paper and (\ref{eq:dCFTcm}) we obtain
\begin{equation}\label{nform}
	\dim^{\text{CFT}}_\BH\text{Higgs}({\cal T}_G(\{Y_{\rm L}, Y_{\rm R}\},\mathfrak{n}))= \mathfrak{n}+\dim(G)+1-d_{Y_{\rm L}}-d_{Y_{\rm R}} ~,
\end{equation}
which is (\ref{eq:dCFTTintro}), reported here for convenience.

\subsection{(Partially) frozen conformal matter theories} \label{sub:pf}

Recall from section \ref{sub:fm5} our definition of frozen conformal matter: it consists in taking to infinity some of the outermost tensor multiplets in the tensor branch description of a conformal matter theory. Recall also that all possible (non-bifurcated) endpoints from \cite{Heckman:2013pva} have a frozen conformal matter representative. 

\begin{figure}[ht]
	\centering
		\includegraphics[height=3in]{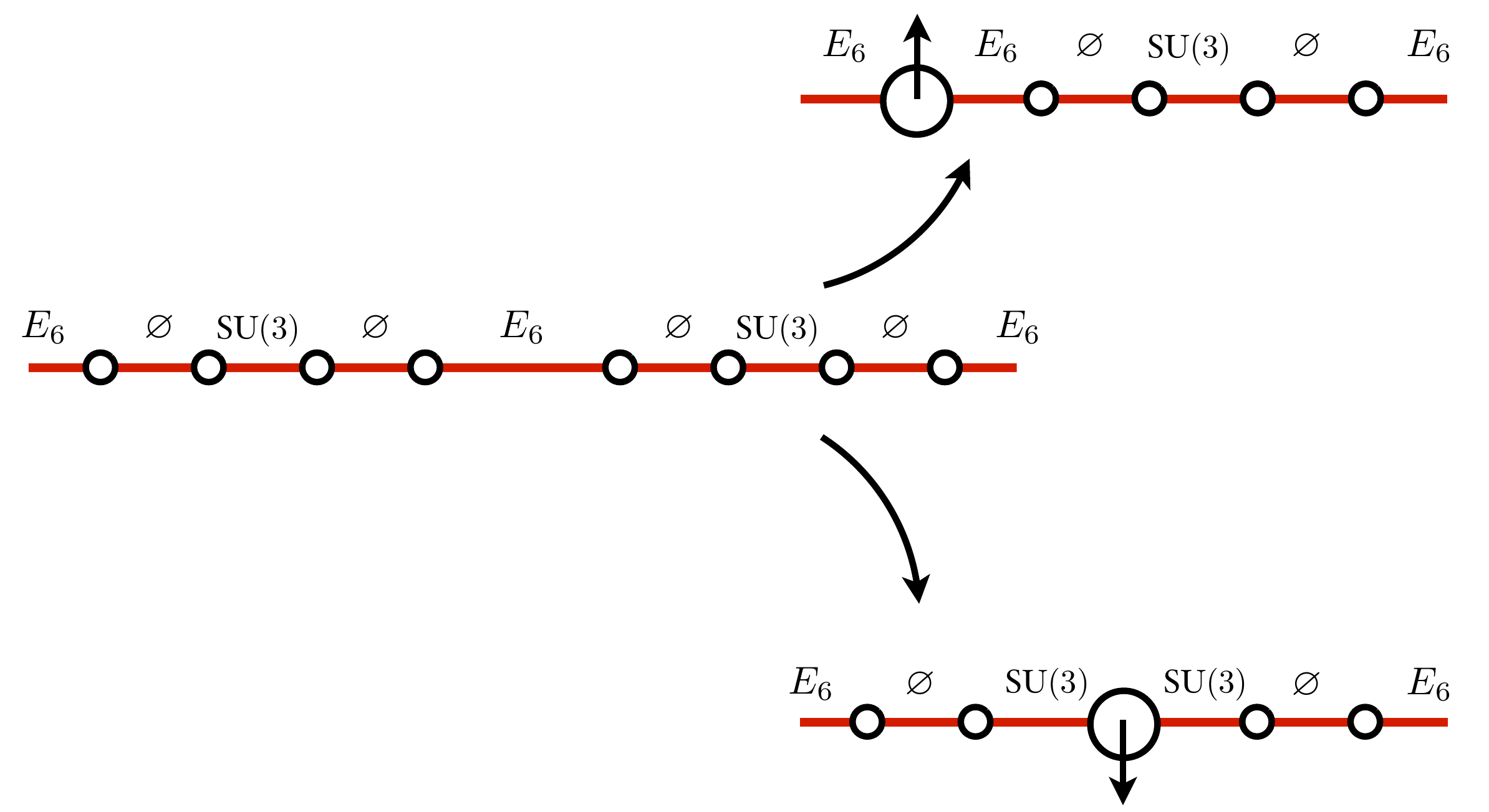}
	\caption{The central part of the picture represents fractional 2 M5-branes (dots) on a $\bR\times \bR^4/\Gamma_{E_6}$ singularity (red line). In this case each of the individual fractions is $1/4$ an ordinary M5. (To be precise, the M5 charges of the fractions are not the same.
	The fraction between $E_6$ and $\varnothing$ has charge $1/3$, while the one between $\varnothing$ and $SU(3)$ has charge $1/6$; see \eref{eq:M5charge}.)  We also show the gauge groups (or lack thereof) on each segment between two fractional M5s. On the top part of the picture, we show a situation where the first four fractions have recombined into a full M5; the latter can now be pulled off the singularity. On the bottom part of the picture we see a different transition, where the fractions have come together in a different way.}
	\label{fig:M5-trans}
\end{figure}

Let us first ask which frozen conformal matter theories are very Higgsable. The list of possible not obviously vH endpoints was obtained in (\ref{eq:vHe}). Using (\ref{eq:end}), (\ref{eq:2-1}), we can see that all the allowed possibilities have a number of curves (before going to the endpoint) which is exactly equal to $\mathfrak{f}(G)-1$, where $\mathfrak{f}$ is the number of fractions (\ref{eq:fr}). For example, for $G=E_6$, these are
\begin{equation}\label{eq:E6vH}
	[1] \,\,\overset{\mathfrak{su}(3)}{3} \,\,  1    \,\,  \overset{\mathfrak{e}_6}{6}\,\, [1] \, ,\qquad
	[SU(3)]\,\,1    \,\,  \overset{\mathfrak{e}_6}{6}\,\, 1 \,\,[SU(3)] ~.
\end{equation}
These are both obtained from the middle part of figure \ref{fig:M5-trans} by keeping four neighboring M5 fractions in the middle and sending to infinity all the others fractions. The same pattern is repeated for any $G$. We will denote these vH theories ${\cal T}^{\rm fr}_{G \to G_{\rm fr}}(0)$, where $G_{\rm fr}$ is the flavor symmetry on either side. For example, (\ref{eq:E6vH}) will be called ${\cal T}^{\rm fr}_{E_6 \to\varnothing}(0)$ and ${\cal T}^{\rm fr}_{E_6\to SU(3)}(0)$.

It is natural to interpret the fact that these theories are very Higgsable as meaning that a transition such as the one in the lower part of figure \ref{fig:M5-trans} can occur. Here we assemble a ``full M5'' from $\mathfrak{f}$ fractions which are not taken in the original order (starting from the unfrozen $G$ singularity), but in any other order. If one takes the full M5 thus formed off the singularity, one leaves behind a partially or totally frozen version of it. In this sense, the theories ${\cal T}^{\rm fr}_{G\to G_{\rm fr}}(0)$ represent an M5 probing a partially frozen singularity.

This conclusion is confirmed if we also look at HN theories. The possible endpoints were obtained in (\ref{eq:HNe}). These now all consist of $N \mathfrak{f}-1$ curves; they are chains of HvH theories such as (\ref{eq:E6vH}), and represent this time $N \mathfrak{f}$ fractions. We will call these theories ${\cal T}^{\rm fr}_{G\to G_{\rm fr}}(N-1)$. They represent $N$ M5-branes probing a partially frozen singularity. The $\mathfrak{n}+1=N$ summand in (\ref{eq:dCFTcm}) represents the moduli of these full M5-branes. (The ${\rm dim}(G)$ summand has to do with the T-brane nilpotent Higgsings, and indeed it gets partially eroded in (\ref{nform}).) If we apply (\ref{eq:dCFTcm}) once again, we get 
\begin{equation}\label{eq:dfcm}
	\dim^{\text{CFT}}_\BH\text{Higgs}({\cal T}_{G\to G_{\rm fr}}(\mathfrak{n}))= \mathfrak{n}+\dim(G_{\rm fr})+1~,
\end{equation}
which is just like (\ref{eq:dCFTcm}), but with $G$ replaced by $G_{\rm fr}$. The Higgs moduli space dimension for these theories should be given by (\ref{nform}) with $G$ replaced by $G_{\rm fr}$ (just as one goes from (\ref{eq:dCFTcm}) to (\ref{eq:dfcm})).

Thus, in the context of frozen conformal matter, HvH and HHN theories represent cases where one assembles M5 fractions into full M5s. We should add that there also exist a few ``shorter'' OvH theories. For instance, $[SU(3)]\,\,  1    \,\,  [E_6]$ is an example of frozen conformal matter in the sense of section \ref{sub:fm5}: it describes two fractional M5s (a half M5) on an $E_6$ singularity. The reason this is very Higgsable has nothing to do with the recombination phenomena of figure \ref{fig:M5-trans}. It is rather a T-brane phenomenon: the Higgs flows activated by the Higgs fields poles (reviewed in section \ref{sub:t}) can also sometimes change the number of tensors. (This does not happen for $G=SU(k)$.) For low $N$ and for nilpotent elements of large enough orbit dimension, this can lead to the loss of all tensor multiplets. We will see other examples of such short OvH frozen conformal matter in section \ref{sub:partial}. 

It would also be possible to Higgs the theories ${\cal T}^{\rm fr}_{G\to G_{\rm fr}}(N)$ by two nilpotent elements $Y_{{\rm L,R}}$ in $G_{\rm fr}$, thus obtaining T-brane theories ${\cal T}^{\rm fr}_{G\to G_{\rm fr}}(\{Y_{\rm L},Y_{\rm R}\},N)$. These theories have not been fully worked out in general, but as we mentioned the cases $G_{\rm fr}=G_2$ and $F_4$ were considered in \cite[Sec.~4.2]{Heckman:2016ssk}.

\subsection{Non-simply-laced $G_{\rm fr}$} 

Notice that $G_{\rm fr}$ is not necessarily simply-laced: looking at (\ref{eq:cm}), we see the non-simply-laced groups $SO(7)$, $USp(2k)$, $G_2$, $F_4$ appearing. In this cases, ${\cal T}^{\rm fr}_{G\to G_{\rm fr}}(N)$ provides a possible definition of $(G_{\rm fr}, G_{\rm fr})$ conformal matter. The $N=0$ cases read
{\small
\be \label{minconfG2}
\begin{split}
[G_2] \bdash [G_2]: &\qquad [G_2] \,\, \overset{\mathfrak{su}_2}2 \,\, 2 \,\, 1\,\, \overset{\mathfrak{e_{8}}}{12}  \,\, 1 \,\,  2 \,\, {\overset{\mathfrak{su_{2}}}2}  \,\,   {\overset{\mathfrak{g_2}}3}  \,\, 1 \,\, \overset{\mathfrak{f_{4}}}5 \,\, 1\,\, [G_2]~, \\
[F_4] \bdash [F_4]: &\qquad  [F_4] \,\,  1 \,\, \overset{\mathfrak{g_{2}}}3 \,\, \overset{\mathfrak{su_{2}}}2 \,\, 2 \,\, 1\,\, \overset{\mathfrak{e_{8}}}{12}  \,\, 1 \,\, 2 \,\,\overset{\mathfrak{su}_2}{2} \,\,\overset{\mathfrak{g}_2}{3} \,\,  1  \,\,   [F_4]~, \\
[Sp(k)] \bdash [Sp(k)] : & \qquad [Sp(k)] \, \, \overset{\mathfrak{so}_{2k+8}}{4} \, \, [Sp(k)]~.
\end{split}
\ee}
We can use these examples to illustrate our general discussion above. If we go to the endpoint locus on the tensor branch of the theories in (\ref{minconfG2}) by blowing down all $(-1)$ curves, we see that the result does not contain only $(-2)$ curves: we get $52$ for the $G_{\rm fr}=G_2$ case, and $4$ for $G_{\rm fr}=F_4$, $Sp(k)$. As we discussed in section \ref{sub:HvHF}, these satisfy the constraint (\ref{eq:main}) for a theory to be hiddenly very-Higgsable. We can also consider longer chains, such as
\be
[F_4] \,\,  1 \,\, \overset{\mathfrak{g_{2}}}3 \,\, \overset{\mathfrak{su_{2}}}2 \,\, 2 \,\, 1\,\, \overset{\mathfrak{e_{8}}}{12}  \,\, 1 \,\, 2 \,\,\overset{\mathfrak{su}_2}{2} \,\,\overset{\mathfrak{g}_2}{3} \,\,  1  \,\,   \overset{\mathfrak{f_{4}}}{5} \,\,  1 \,\, \overset{\mathfrak{g_{2}}}3 \,\, \overset{\mathfrak{su_{2}}}2 \,\, 2 \,\, 1\,\, \overset{\mathfrak{e_{8}}}{12}  \,\, 1 \,\, 2 \,\,\overset{\mathfrak{su}_2}{2} \,\,\overset{\mathfrak{g}_2}{3} \,\,  1  \,\, [F_4]~.
\ee
The endpoint is now $33$. This indeed appears in the list of possibilities (\ref{eq:HNe}) for $\mathfrak{n}=1$. One can check this directly as well: $\eta^{\rm end} = \left(\begin{smallmatrix} 3 & -1 \\ -1 & 3 \end{smallmatrix}\right)$, $\eta_{\rm end}^{-1} = \frac18\left(\begin{smallmatrix} 3 & 1 \\ 1 & 3 \end{smallmatrix}\right)$, $n^{\text{end}}_T= 2$; thus $\mathfrak{n} = 2-1=1$.

\subsection{Complete freezing: $G_{\rm fr}=\varnothing$} 

It is also possible that $G_{\rm fr}=\varnothing$. We already saw one such case in the first theory in (\ref{eq:E6vH}). The other cases read 
\be \label{frozenmin}
\begin{split}
G=SO(8): &\qquad [1] \,\,\overset{\mathfrak{so}(8)}{4}\,\, [1]~, \\
G=E_6: &\qquad [1] \,\,\overset{\mathfrak{su}(3)}{3} \,\,  1    \,\,  \overset{\mathfrak{e}_6}{6}\,\, [1]~, \\
G= E_7: &\qquad  [1] \,\,\overset{\mathfrak{su}(2)}{2} \,\,\overset{\mathfrak{so}(7)}{3} \,\,\overset{\mathfrak{su}(2)}{2} \,\,  1  \,\, \overset{\mathfrak{e}_7}{8} \,\, [1]~, \\
G= E_8: &\qquad  [1]\,\, 2 \,\,\overset{\mathfrak{su}(2)}{2} \,\,\overset{\mathfrak{g}_2}{3} \,\,  1  \,\, \overset{\mathfrak{f}_4}{5} \,\, 1 \,\, \overset{\mathfrak{g}_2}{3}\,\,\overset{\mathfrak{su}(2)}{2} \,\, 2 \,\, 1  \,\,  \overset{\mathfrak{e}_8}{12}\,\, [1]~.
\end{split}
\ee
As we mentioned, we call these theories ${\cal T}^{\rm fr}_{G \to G_{\rm fr}}(0)$. Notice, however, that for $G=E_8$ there are two more possibilities to have $G_{\rm fr}= \varnothing$:
\begin{equation}\label{eq:ex-fr}
\begin{split}
	&[2] \,\,\overset{\mathfrak{su}(2)}{2} \,\,\overset{\mathfrak{g}_2}{3} \,\,  1  \,\, \overset{\mathfrak{f}_4}{5} \,\, 1 \,\, \overset{\mathfrak{g}_2}{3}\,\,\overset{\mathfrak{su}(2)}{2} \,\, 2 \,\, 1  \,\,  \overset{\mathfrak{e}_8}{12}\,\, 1\,\,[2]~,\\
	&[1]  \,\, \overset{\mathfrak{f}_4}{5} \,\, 1 \,\, \overset{\mathfrak{g}_2}{3}\,\,\overset{\mathfrak{su}(2)}{2} \,\, 2 \,\, 1  \,\,  \overset{\mathfrak{e}_8}{12}\,\, 1\,\,2\,\,\overset{\mathfrak{su}(2)}{2} \,\,\overset{\mathfrak{g}_2}{3} \,\,[1]~.
\end{split}
\end{equation}
We will call these two ``exotically frozen''\footnote{The term {\it exotic} is used here because, as we shall discuss in section \ref{sec:anomfrozen}, the theories \eref{eq:ex-fr} have more complicated anomaly formulae than their non-exotic counterpart.}.   Notice that they correspond to the case $d_{E_8\to \varnothing}=5$ in table \ref{tab:d}, whereas the case $G=E_8$ in (\ref{frozenmin}) corresponds to $d_{E_8\to \varnothing}=6$ (see \cite[Eq.~(3.6)]{Ohmori:2015pua}). 

Up to reflection, (\ref{frozenmin}) and (\ref{eq:ex-fr}) exhaust all the possible cases where $G_{\rm fr}=\varnothing$. 

\subsection{Partial recombinations} 
\label{sub:partial}

Our methods allow us to access the Higgs moduli space at the origin of the tensor branch for several interesting theories; one might also wonder, however, about the Higgs moduli space on different non-generic loci of the tensor branch. For example, let us now go back to the original unfrozen conformal matter chains ${\cal T}_G(N)$. We know that the Higgs moduli space has dimension (\ref{eq:dCFTcm}) at the CFT point, where all the fractions are coinciding. But what about loci where only \textit{some} of the fractions are coinciding?

Let us go back to (\ref{eq:dH}) and rewrite it as $d_H = 29(n_T- \mathfrak{n}) + n_H-n_V$. In terms of the number of fractions $\mathfrak{f}$ (recall (\ref{eq:fr})), this is $29 (\mathfrak{f}-1)\mathfrak{n} + n_H-n_V$. On a generic point of the tensor branch, the physics is weakly coupled and one expects simply $n_H-n_V$. Thus one expects that the moduli space increases every time one puts fractions on top of each other, up to a maximum which is reached one has made  $(\mathfrak{f}-1)\mathfrak{n}$ coincidences; this is not the same as making all the M5s coincide, but rather the same as recombining the fractions in several full M5s. 

In other words, the dimension (\ref{eq:dCFTcm}) is also valid at points where the M5 fractions are not all on top of each other, but also in loci of the tensor branch where the fractions coincide in groups of $\mathfrak{f}$, thus making full M5s which need not themselves be on top of each other. To check this, let us point out that if we have a sequence of CFT's connected by vectors and hypers, we expect $d_H$ at that locus to be 
\begin{equation}\label{eq:partial}
	\sum_i d_H({\rm CFT}_i)+ n_H-n_V ~.
\end{equation}
The non-generic locus of ${\cal T}_G(N-1)$ where $N$ full M5s have formed but do not all coincide can be thought of as $N$ copies of ${\cal T}_G(0)$ connected by $(N-1)$ copies of a $G$ gauge field. Thus (\ref{eq:partial}) gives 
\begin{equation}
	({\rm dim}(G)+1)N - {\rm dim}(G)(N-1) = {\rm dim}(G)+ N\,
\end{equation}
which coincides with (\ref{eq:dCFTcm}), recalling $N=\mathfrak{n}+1$. Thus the maximum Higgs moduli space dimension is already reached on this locus. 

On the other hand, recall we mentioned the existence of some ``short'' vH theories that are partially frozen conformal matter in the sense of section \ref{sub:fm5} (namely, they are incomplete conformal matter chains), but that do not consist of $\mathfrak{f}$ fractions. There are some obvious examples, such as a single $-1$-curve, but also a few less-obvious ones, such as
\be \label{eq:short}
\begin{split}
[E_7] \rdash [SO(7)]: &\quad [E_7] \,\, {1} \,\,\overset{\mathfrak{su}(2)}{2} \,\,[SO(7)]~, \\
[E_8] \rdash [G_2]:&\quad   [E_8] \,\, {1}\,\, 2 \,\,\overset{\mathfrak{su}(2)}{2} \,\,[G_2]~, \\
[E_8] \rdash [F_4]: &\quad [E_8] \,\, {1}\,\, 2 \,\,\overset{\mathfrak{su}(2)}{2} \,\,\overset{\mathfrak{g}_2}{3} \,\,  1  \,\, [F_4]~.
\end{split}
\ee
These are all OvH theories, as one can easily check. The first describes $3=\frac12\mathfrak{f}(E_7)$ fractions, which means it is a ``half M5'' on an $E_7$ singularity; the second and third describe one third and one half an M5 on top of an $E_8$ singularity. 

Using these theories, we can access the Higgs moduli space at some loci of partial coincidence of fractions. For example, for $G=E_7$, we can consider the locus where the fractions are coinciding 3 at a time, so that we have a sequence of $2N$ half-M5s:
\begin{equation}
	[E_7]\rdash (SO(7))\rdash (E_7) \cdots (E_7) \rdash (SO(7))\rdash [E_7]~.
\end{equation}
Since $[E_7] \rdash [SO(7)]$ is OvH, we can use (\ref{eq:dCFT}) to compute $d_H=8-3+2\cdot 29=63$. Now (\ref{eq:partial}) gives $(2N \times 63)-21 N-133(N-1)=133 - 28 N = {\rm dim}(E_7)-28N$. Comparing this with (\ref{eq:dCFTcm}), we see that we have the Higgs moduli space dimension has gone down by $29 N$ --- namely, we lose a dimension 29 whenever we split an M5 in two halves. 

In the same vein we can consider $[E_8]\rdash (F_4)\rdash (E_8) \cdots \rdash[E_8]$, which is a sequence of half-M5s on an $E_8$ singularity. Here $[E_8]\rdash [F_4]$ has $d_H=136$, and  (\ref{eq:partial}) gives $(2N \times 136)-52 N-248(N-1)=248 - 28 N = {\rm dim}(E_8)-28N$. Again we have lost a dimension 29 by splitting the M5s in half. A similar computation can be performed for an $E_6$ singularity, with the same conclusion.

\subsection{Anomaly polynomials of frozen conformal matter} \label{sec:anomfrozen}

We will now compute the anomaly polynomial for $(G_{\mathrm{fr}},G_{\mathrm{fr}})$ conformal matter in a similar fashion to (3.19), (3.23) and (B.20) of \cite{Ohmori:2014kda}, for a chain ${\cal T}^{\rm fr}_{G\to G_{\mathrm{fr}}}(Q-1)$ of $Q$ copies of $(G_{\mathrm{fr}},G_{\mathrm{fr}})$ conformal matter theories:
\be \label{chainfrozen}
[G_{\mathrm{fr}}] \bdash (G_{\mathrm{fr}}) \bdash  \cdots \bdash (G_{\mathrm{fr}}) \bdash  [G_{\mathrm{fr}}]~.
\ee
$G_{\mathrm{fr}}$ can also be non-simply-laced or trivial.

We find that the anomaly polynomial, including the center of mass tensor multiplet, admits an elegant expression which is a simple generalisation of (B.20) of \cite{Ohmori:2014kda}:
\be
I^{\text{tot}} = \frac{1}{24} Q^3 |\Gamma_G|^2 c_2(R)^2 - Q I_8 - \frac{1}{2} Q |\Gamma_G| (J_{4,L}+ J_{4,R}) -\frac{1}{2} I^{\text{vec}}_L -\frac{1}{2} I^{\text{vec}}_R~,
\ee
where $G$ is the simply-laced group from which $G_{\mathrm{fr}}$ is originated, and other definitions are as follows:
\bi
\item The expression for $I_8$ is
\be
I_8 = \frac{1}{48} \left[ p_2(T) - p_1(T) c_2(R) - \frac{1}{4} p_1(T)^2 \right]~.
\ee
\item The expression for $J_{4,L/R}$ is
\be \label{eq:J4LR}
J_{4,L/R} = \frac{1}{48}  (4 c_2(R) + p_1(T)) \chi_{G \rightarrow G_\text{fr}} + \frac{1}{4 d_{G\rightarrow G_\text{fr}}} \tr F^2_{L/R}~,  
\ee
where
\begin{equation}\label{eq:chinice}
	\chi_{G\to G_{\mathrm{fr}}} = r_G-11+\frac{12}{d_{G\to G_{\mathrm{fr}}}} -\frac1{|\Gamma_G|}~.
\end{equation}
$r_G$ is the rank of $G$, $|\Gamma_G|$ is the cardinality of $\Gamma_G$, and $d_{G \rightarrow G_{\text{fr}}}$ is the parameter given in (\ref{tab:d}). When $G_{\mathrm{fr}}=G$, $d_{G\to G_{\mathrm{fr}}}$ is taken to be 1. The last term in (\ref{eq:J4LR}) is present only when $G_{\text{fr}}$ is non-empty.
The quantity $\chi_{G\to G_{\mathrm{fr}}}$ can also be expressed using the ranks of $G_{\mathrm{fr}}$ as
\be \label{eq:chiGfr}
 \chi_{G  \rightarrow G_\text{fr}} = 
 \begin{cases} r_{G}+1-\frac{1}{|\Gamma_G|}  &\qquad \text{unfrozen, $G_\text{fr} = G$}~; \\
 -\frac{3}{5} -\frac{1}{|\Gamma_G|}  &\qquad \text{``exotically frozen'': (\ref{eq:ex-fr})}~; \\
 r_{G_\text{fr}} - 1  -\frac{1}{|\Gamma_G|} &\qquad \text{otherwise}~.
\end{cases} 
\ee

\item The expression for $I^{\text{vec}}_{L/R}$ is
\be
\begin{split}
I^{\text{vec}}_{L/R} &= -\frac{1}{24} \left[ \tr_{\text{adj}}F^4_{L/R} + 6 c_2(R) \tr_{\text{adj}}F^2_{L/R}+ \dim(G_{\text{fr}}) c_2(R)^2 \right]  \\
& \qquad - \frac{1}{48} p_1(T) \left[ \tr_{\text{adj}}F^2_{L/R} +  \dim(G_{\text{fr}})  c_2(R) \right] - \frac{1}{5760} \left[7 p_1(T)^2 - 4 p_2(T) \right]~,
\end{split}
\ee
where 
\be
 \tr_{\text{adj}}F^2_{L/R} = h^\vee(G_{\text{fr}}) \tr F^2_{L/R}~.
\ee
The latter are of course only present when $G_{\rm fr}\neq \varnothing.$ 
\ei

The center of mass contribution to the above anomaly polynomial is similar to (3.24) of \cite{Ohmori:2014kda}:
\be
\begin{split}
I^{\text{CM}} &= \left[  \frac{1}{24} c_2(R)^2+\frac{1}{48} c_2(R) p_1+\frac{23 }{5760}p_1(T)^2-\frac{116}{5760} p_2(T) \right] \\ 
& \qquad + \frac{1}{2Q} \frac{1}{16 d^2_{G \rightarrow G_\text{fr}}} \left(\tr F_L^2 - \tr F_R^2   \right)^2~.
\end{split}
\ee
In particular, the second line is a Green--Schwarz contribution.

The anomaly polynomial of the SCFT associated with \eref{chainfrozen}, which is equal to $I^{\text{tot}} - I^{\text{CM}}$, can therefore be written as
\be
\begin{split}\label{eq:anomcoeff}
&\alpha c_2(R)^2+ \beta c_2(R) p_1(T) + \gamma p_1(T)^2 + \delta p_2(T)  \\
& + \left( -\frac{x}{8} c_2(R) +\frac{h^\vee_{G_{\text{fr}}}}{96} p_1(T) \right) (\tr F_L^2+\tr F_R^2)  \\
&+ \frac{1}{48} (\tr_{\text{adj}} F_L^4 + \tr_{\text{adj}} F_R^4) -\frac{1}{2Q}\frac{1}{16 d^2_{G \rightarrow G_\text{fr}}}  \left(\tr F_L^2 - \tr F_R^2   \right)^2 ~
\end{split}
\ee
where
\be 
\begin{split}
\alpha &= \frac{1}{24}|\Gamma_G|^2 Q^3 - \frac{1}{12} Q |\Gamma_G|  \chi_{G  \rightarrow G_\text{fr}} + \frac{1}{24} (\mathrm{dim}(G_\text{fr}) - 1) \\
\beta &= \frac{1}{48} Q \left(1- |\Gamma_G|  \chi_{G  \rightarrow G_\text{fr}} \right) + \frac{1}{48} (\mathrm{dim}(G_\text{fr}) - 1) \\
\gamma &= \frac{1}{5760} \left[ 30(Q-1) + 7( \mathrm{dim}(G_\text{fr}) + 1) \right]  \\
\delta &= -\frac{1}{1440} \left[ 30(Q-1) + \mathrm{dim}(G_\text{fr}) + 1 \right] \\
x &=   \frac{|\Gamma_{G}|}{d_{G \rightarrow G_\text{fr}}} Q - h^\vee_{G_\text{fr}} ~.
\end{split}
\ee

A special case that is worth mentioning here is when $G_{\text{fr}} = \varnothing$ and $G_{\text{fr}}$ is not exotically frozen. Using (\ref{eq:chinice}) or (\ref{eq:chiGfr}), we take $ \chi_{G  \rightarrow G_\text{fr}} = - 1  -\frac{1}{|\Gamma_G|}$ and we obtain
\be
\begin{split}
\alpha &= \frac{1}{24} |{\Gamma}_G|^2 Q^3 - \frac{1}{12} ( -|{\Gamma}_G|- 1) Q- \frac{1}{24} \\
\beta &= \frac{Q}{48}(2+|{\Gamma}_G|) -\frac{1}{48} \\
\gamma &= \frac{1}{5760} [30 (Q-1)+7] \\
\delta &= -\frac{1}{1440}[ 30 (Q-1)+1]~.
\end{split}
\ee
This can be obtained from (3.19) and (3.23) of \cite{Ohmori:2014kda} with $\dim G=\mathrm{rank} \, G  =0$ and with $|\Gamma_G|$ replaced by $- |\Gamma_G|$.

Our formulas for the anomaly polynomial might give interesting indications on the physics of frozen singularities. Notice in particular that our results are a minimal modification of those in \cite{Ohmori:2014kda}: the new elements are the appearance of the parameter $d_{G \to G_{\rm fr}}$ of (\ref{tab:d}), and of $\chi_{G\to G_{\rm fr}}$ in (\ref{eq:chiGfr}). It would be very interesting to see how this comes about from an anomaly inflow computation similar to that in App.~B of \cite{Ohmori:2014kda}, where for example our $\chi_{G\to G_{\rm fr}}$ appears to modify the $\chi_\Gamma$ of their (B.7).

\section{Conformal matter theories on $T^2$}
\label{sec:sl-cm}

We will now start comparing the results we have obtained so far with various existing results for torus compactifications. We expect that the Higgs branch dimension of the original 6d theory matches with that of its $T^2$ compactification, and we will check this for various examples. Moreover, using \eref{eq:dCFT} and \eref{nform} as a tool, from the next section we will propose new results for the $T^2$ and $T^3$ compactifications of HvH and HHN theories, including $(G,G)$ conformal matter theories with $G$ non-simply-laced and worldvolume theories of M5-branes probing the completely frozen singularity.

We will start in this section with  $(G,G)$ conformal matter; as we have seen, this is the theory describing $N$ M5-branes on a $\bR\times \bC^2/\Gamma_G$ singularity. We have seen its Higgs moduli space is given by (\ref{eq:dCFTcm}), with $N=\mathfrak{n}+1$. We will provide some checks of (\ref{eq:dCFTcm}) against the known results in  \cite{Ohmori:2015pua, Ohmori:2015pia}.

\paragraph{The case $N=1$.} For $N=1$ (or $\mathfrak{n}=0$), the 6d theory in question, ${\cal T}_G(0)$, is a minimal conformal matter of type $(G,G)$. The $T^2$ compactification of the CFT point of such a theory \cite{Ohmori:2015pua} is a class $\CS$ theory of type $G$ whose Gaiotto curve is a sphere with two maximal punctures (those associated with the trivial nilpotent orbit) and one minimal punctures (associated with the so-called subregular orbit). In the notation of  \cite{Ohmori:2015pua} this reads\footnote{In this paper, we denote by $\mathsf{S} \langle \Sigma \rangle_G\{ Y_1, Y_2, \ldots, Y_\ell \}$ a four-dimensional class $\CS$ theory of type $G$ whose Gaiotto curve is a Riemann surface $\Sigma$ with punctures $Y_1, Y_2, \ldots, Y_\ell$.}
\begin{equation}\label{eq:S00O}
	\mathsf{S} \langle S^2 \rangle_G\{ 0, 0, \varnothing \}~.
\end{equation}
 Recall that the Higgs branch dimension for a class $\CS$ theory of type $G$ and punctures ${\cal O}_i$  is given by
\begin{equation}\label{eq:SH}
	{\rm dim}_\BH \text{Higgs}(\mathsf{S} \langle S^2 \rangle_{G}\{ 0, Y_{\rm L}, Y_{\rm R} \})=\frac32\left(\dim(G)- \mathrm{rank}(G) \right)  - {\rm dim}({\cal O}_i)+ \mathrm{rank}(G)\, .
\end{equation}
Moreover, the subregular orbit has dimension $\frac{1}{2}\left(\dim(G)- \mathrm{rank}(G) \right)-1 $.
The Higgs branch dimension of (\ref{eq:S00O}) is then
\be
\begin{split}
&\frac{3}{2}\left(\dim(G)- \mathrm{rank}(G) \right) - 0 -0 - \left[ \frac{1}{2}\left(\dim(G)- \mathrm{rank}(G) \right)-1 \right] + \mathrm{rank}(G) \\
&=1+\dim(G)~,
\end{split}
\ee
in agreement with \eref{eq:dCFTcm}.

\paragraph{$G=SU(k)$ and general $N$.}  Here $\Gamma = \BZ_k$.  For $N> k$, the $T^2$ compactification of the 6d theory ${\cal T}_{SU(k)}(N-1)$ is given by \cite[Eq.~(5.10),(5.11)]{Ohmori:2015pia}:\footnote{According to (4.17) of \cite{Ohmori:2015pia}, the case of $k< N$ can be obtained by exchanging $k$ and $N$, and the case of $k=N$ can be obtained by replacing $ \mathsf{S} \langle S^2 \rangle_{SU(N)}\{[1^N],[1^N], [N-k,1^k]  \}$ by  $\mathsf{S} \langle S^2 \rangle_{SU(N)}\{[1^N],[1^N], [1^N]  \}$ with one flavour of the hypermultiplet in the fundamental representation of $SU(N)$.}
\be \label{compTconfmatt}
\frac{ \mathsf{S} \langle S^2 \rangle_{SU(N)}\{[1^N],[1^N], [N-k,1^k]  \} \times \mathsf{S} \langle S^2 \rangle_{SU(k)}\{ [1^{k}], [1^k] ,[1^k] ] \}}{SU(N) \times \diag(SU(k) \times SU(k))}
\ee
where $SU(N)$ in the denominator denotes the gauging of the diagonal $SU(N)$ subgroup of the flavour symmetry $SU(N) \times SU(N)$ coming from the two copies of $[1^N]$, and $\diag(SU(k) \times SU(k))$ in the denominator denotes the gauging of diagonal $SU(k)$ subgroup of the flavour symmetry $SU(k) \times SU(k)$ coming from $[N-k,1^k]$ and one of the $[1^k]$ punctures.  The former gauging can be regarded as forming a handle so that the resulting theory becomes a torus; see (4.17) of \cite{Ohmori:2015pia}.  Thus, at a generic point of the Higgs branch, there is an unbroken $U(1)^{N-1}$ gauge symmetry (see, \eg~ \cite{Hanany:2010qu}).   The Higgs branch dimension of \eref{compTconfmatt} is therefore, recalling (\ref{eq:SH}),
\be \label{dimHiggscompTconfmatt}
\begin{split}
\dim_{\BH} \text{Higgs of \eref{compTconfmatt}} &= \left[ \frac{3}{2}(N^2-N) - d_{[N-k,1^k]} +(N-1) \right] \\
& \qquad + \left[ \frac{3}{2}(k^2-k)  +(k-1) \right] \\
& \qquad  -(N^2-1)-(k^2-1) +(N-1) \\
&=  N +(k^2-1)~.
\end{split}
\ee
where the last term $(N-1)$ in the first equality corresponds to the unbroken $U(1)^{N-1}$ gauge symmetry.  We have used that the dimension of the orbit $[N-k, 1^k]$ is
\be
d_{[N-k,1^k]} = \frac{1}{2} (N - k-1 ) (k + N)~.
\ee
We see that \eref{dimHiggscompTconfmatt} is in agreement with \eref{eq:dCFTcm}.  

\paragraph{The case of $G=SO(2k)$ and general $N$.}  The $T^2$ compactification of the conformal matter theory in question is described by \cite[Eq.~(3.3.57)]{Ohmorithesis2015}:
\be \label{confmattDkn}
\frac{\mathsf{S} \langle S^2 \rangle_{SU(2N)}\{[2^N], \underline{TM}, \underline{O}_k \} \times \mathsf{S} \langle S^2 \rangle_{SO(2k)}\{ [1^{2k}], [1^{2k}], [1^{2k}] ] \}}{SU(N) \times \diag(SO(2k) \times SO(2k))}
\ee
where $\mathsf{S} \langle S^2 \rangle_{SU(2N)}\{[2^N], \underline{TM}, \underline{O}_k \}$ denotes a 4d class $\CS$ theory of $SU(2N)$ type, whose Gaiotto curve is a sphere with a untwisted $[2^N]$ puncture, a minimal twisted puncture $\underline{TM}$, and a twisted puncture $\underline{O}_k$ that has an $SO(2k)$ flavour symmetry.  Using the notation of \cite{Chacaltana:2012ch} for the twisted $SU(2N)=A_{2N-1}$ class $\CS$ theory, the twisted puncture is labelled by a B-partition of $2N+1$, and the untwisted puncture is labelled by an ordinary partition of $2N$.  The minimal twisted puncture $\underline{TM}$, whose flavour symmetry is empty, is labelled by $[2N+1]$.  The twisted puncture of $\underline{O}_k$, whose flavour symmetry is $SO(2k)$, is labelled by $[2(N-k)+1,1^{2k}]$.  Here, the $SU(N)$ factor in the denominator denotes the gauging of the flavour symmetry $SU(N)$ associated with $[2^N]$, and $\diag(SO(2k) \times SO(2k))$ denotes the gauging of the diagonal subgroup of $SO(2k) \times SO(2k)$ coming from $\underline{O}_k$ and one of the $[1^{2k}]$ punctures.  Similarly to the previous case, the $SU(N)$ gauging forms a handle; see \cite[Eq.~(3.3.57)]{Ohmorithesis2015}.

Let us now compute the Higgs branch of \eref{confmattDkn}. We first compute the Higgs branch dimension of each component:
\be
\dim_{\BH}  \mathsf{S} \langle S^2 \rangle_{SO(2k)}\{ [1^{2k},1^{2k},1^{2k} ] \}= k+\left(3 \times 2\sum_{j=1}^{k-1} j\right) = k(3k-2) ~,
\ee
and
\be \label{classSSU2Nwithtwist}
\begin{split}
&\dim_{\BH} \mathsf{S} \langle S^2 \rangle_{SU(2N)}\{[2^N], \underline{TM},\underline{O}_k \} \\
& = \left( \frac{1}{2} \left(2 N^2-N+1\right) - d_{\underline{TM}} \right) + \left( \frac{1}{2} \left(2 N^2-N+1\right) - d_{\underline{O}_k} \right) \\
& \qquad + \left( \frac{1}{2}[ \{ (2N)^2-1\} -(2N-1)] -d_{[2^N]} \right) +(2N-1) \\
&  =  \left( \frac{1}{2} \left(2 N^2-N+1\right) -  N^2 \right) + \left( \frac{1}{2} \left(2 N^2-N+1\right) - (N^2-k^2) \right) \\
&  \qquad + \left( \frac{1}{2}[ \{ (2N)^2-1\} -(2N-1)] -N^2 \right) +(2N-1) \\
&  = N^2+k^2 \\
\end{split}
\ee
where $\frac{1}{2} \left(2 N^2-N+1\right)$ is the value of $n_H-n_V$ for the trivial twisted orbit of $SO(2N+1)$; see (3.46) of \cite{Chacaltana:2012zy}.  Thus, the Higgs branch dimension of \eref{confmattDkn} is given by
\be
\begin{split}
\dim_{\BH} \text{Higgs of \eref{confmattDkn}} &= k(3k-2) +(N^2+k^2) \\
& \qquad  -(N^2-1)-\frac{1}{2}(2k)(2k-1) + (N-1)\\
&=  N +\frac{1}{2}(2k)(2k-1)~,
\end{split}
\ee
in agreement with \eref{eq:dCFTcm}.  Note that the terms $(N-1)$ in the second line comes from the fact that $SU(N)$ gauge symmetry is broken to $U(1)^{N-1}$ at a the generic point on the Higgs branch.

It was also pointed out in \cite[Eq.~(3.3.58)]{Ohmorithesis2015} that the $SU(N)$ gauge field in \eref{confmattDkn} can be absorbed in to the twisted class $\CS$ theory and this results in the following description:
\be \label{confmattDknII}
\frac{\mathsf{S} \langle T^2 \rangle_{SU(2N)}\{\underline{TM},\underline{TM},\underline{TM}, \underline{O}_k \} \times \mathsf{S} \langle S^2 \rangle_{SO(2k)}\{ [1^{2k}], [1^{2k}], [1^{2k}] ] \}}{\diag(SO(2k) \times SO(2k))}
\ee
The Higgs branch dimension of $\mathsf{S} \langle T^2 \rangle_{SU(2N)}\{\underline{TM},\underline{TM},\underline{TM}, \underline{O}_k \}$ can be similarly computed as the above:\footnote{The mirror of the three dimensional theory coming from $S^1$ compactification of $\mathsf{S} \langle T^2 \rangle_{SU(2N)}\{\underline{TM},\underline{TM},\underline{TM}, \underline{O}_k \}$ is $S(USp(2N))-(O(2k))-(USp(2k-2))-(O(2k-2)) - \cdots -(USp(2))-(O(2))$, where $S$ denotes an adjoint hypermultiplet of $USp(2N)$ gauge group \cite{Benini:2010uu, Cremonesi:2014vla}. The Coulomb branch dimension of this mirror theory is $2\left( \sum_{j=1}^{k-1} j \right) +k +N = k^2+N$.  This is in agreement with the above Higgs branch dimension.}
\be 
\begin{split}
&\dim_{\BH} \, \text{Higgs of $\mathsf{S} \langle T^2 \rangle_{SU(2N)}\{\underline{TM},\underline{TM},\underline{TM}, \underline{O}_k \}$} \\
&=   3 \left( \frac{1}{2} \left(2 N^2-N+1\right) -  N^2 \right) + \left( \frac{1}{2} \left(2 N^2-N+1\right) - (N^2-k^2) \right)  \\
& \qquad +(2N-1) + (N-1)\\
&= k^2+N~.
\end{split}
\ee
Thus, the Higgs branch dimension of \eref{confmattDknII} is 
\be
\begin{split}
\dim_{\BH} \, \text{Higgs of \eref{confmattDknII}} &= (k^2+N) +  k(3k-2)  -\frac{1}{2}(2k)(2k-1)  \\
&= N +\frac{1}{2}(2k)(2k-1)~,
\end{split}
\ee
in agreement with \eref{eq:dCFTcm}.

In the next section, we generalize these results to more general T-brane theories.
\section{T-brane theories on $T^2$}
\label{sec:T}

As we saw in section \ref{sub:t}, one can Higgs the $G \times G$ flavor symmetry in the conformal matter theory ${\cal T}_G(N)$ of type $(G,G)$ by the nilpotent orbits $(Y_{\rm L}, Y_{\rm R})$, obtaining the so-called T-brane theory $\CT_{G} ( \{Y_{\rm L}, Y_{\rm R}\} , N)$. 

We claim the $T^2$ compactification of $\CT_G ( \{Y_{\rm L}, Y_{\rm R}\} ,N-1 )$ is a simple generalisation of (4.17) and (5.10) of \cite{Ohmori:2015pia}:
\be \label{T2compTbrane}
\frac{\CS^{4d}_{(\varnothing, G)} \{SU(N), G \}  \times \mathsf{S} \langle S^2 \rangle_{G}\{ 0, Y_{\rm L}, Y_{\rm R} \}}{SU(N) \times \diag(G \times G)}~.
\ee
The notation is as follows:
\bi
\item $\mathsf{S} \langle S^2 \rangle_{G}\{ 0, Y_{\rm L}, Y_{\rm R} \}$ denotes the class $\CS$ theory of type $G$ associated with a sphere with one maximal puncture $0$, and two other punctures $Y_{\rm L}$ and $Y_{\rm R}$.
\item $\CS^{4d}_{(\varnothing, G)} \{SU(N), G \}$ is a certain 4d field theory with a flavour symmetry $SU(N) \times G$.  Upon gauging the $SU(N)$ symmetry, the theory $\CS^{4d}_{(\varnothing, G)} \{SU(N), G \}/SU(N)$ can be realised as the $T^2$ compactification of  \cite[Sec.~5.4]{Ohmori:2015pia}
\be \label{6doriginS4d}
\CT_G(\{ \varnothing, (0)\}, \mathfrak{n}' ) \underbrace{\bdash (G) \bdash (G) \ldots (G)\bdash}_{N-\mathfrak{n}'-1~\text{copies of $\bdash$ }}  [G]~.
\ee
$\CT_G(\{ \varnothing, (0) \}, \mathfrak{n}')$ denotes  the T-brane theory \cite{Heckman:2016ssk} of type $(G,G)$ associated with the principal orbit $\varnothing$ and the trivial orbit (\ie~maximal puncture) $(0)$, with the flavour symmetry $G$ of the latter being gauged. As introduced in (\ref{eq:bdash}), the notation $[G] \bdash [G]$ denotes the minimal conformal matter of type $(G,G)$. We will make (\ref{6doriginS4d}) more explicit shortly.

For $G=SU(k)$ or $G=SO(2k)$, $\CS^{4d}_{(\varnothing, G)} \{SU(N), G \}$ is in fact a class $\CS$ theory discussed in the previous section:
\be \label{knowncompS4d}
\begin{split}
\CS^{4d}_{(\varnothing, SU(k))} \{SU(N), SU(k) \}  &= \mathsf{S} \langle S^2 \rangle_{SU(N)}\{[1^N],[1^N], [N-k,1^k]  \} \\
\CS^{4d}_{(\varnothing, SO(2k))} \{SU(N), SO(2k) \}  &=\mathsf{S} \langle S^2 \rangle_{SU(2N)}\{[2^N], \underline{TM}, \underline{O}_k \} ~, 
\end{split}
\ee
where the notation in the latter equation is explained around \eref{confmattDkn}.  The description in terms of a class $\CS$ theory for $G=E_{6,7,8}$ is currently not known.
\item The $SU(N)$ factor in the denominator denotes the gauging of the flavour symmetry $SU(N)$ in $\CS^{4d}_{(\varnothing, G)} \{SU(N), G \}$, and the factor $\diag(G \times G)$ denotes the gauging of the diagonal subgroup of $G \times G$ coming from $\CS^{4d}_{(\varnothing, G)} \{SU(N), G \} $ and the puncture $0$ in $\mathsf{S} \langle S^2 \rangle_{G}\{ 0, Y_{\rm L}, Y_{\rm R} \}$.
\ei

Using \eref{T2compTbrane} and \eref{knowncompS4d}, we can compute the Higgs branch in a similar way as in \eref{dimHiggscompTconfmatt} and \eref{classSSU2Nwithtwist}.  For $G=SU(k)$, the result is
\be
\begin{split}
\dim_{\BH} \text{Higgs of \eref{T2compTbrane}} &= \left[ \frac{3}{2}(N^2-N) - d_{[N-k,1^k]} +(N-1) \right] \\
& \qquad + \left[ \frac{3}{2}(k^2-k)  -d_{Y_{\rm L}} - d_{Y_{\rm R}}+(k-1) \right] \\
& \qquad  -(N^2-1)-(k^2-1) +(N-1) \\
&=  N +(k^2-1) -d_{Y_{\rm L}} - d_{Y_{\rm R}} ~,
\end{split}
\ee
in agreement with \eref{nform} upon setting $\mathfrak{n}=N-1$. For $G=SO(2k)$, the result is
\be
\begin{split}
\dim_{\BH} \text{Higgs of \eref{T2compTbrane}} &= (N^2+N+k^2-1) + \left( 3 k(k-1) - d_{Y_{\rm L}} - d_{Y_{\rm R}} +k \right) \\
& \qquad  -(N^2-1)-\frac{1}{2}(2k)(2k-1) \\
&=  N +\frac{1}{2}(2k)(2k-1) - d_{Y_{\rm L}} - d_{Y_{\rm R}}~,
\end{split}
\ee
in agreement with \eref{nform} upon setting $\mathfrak{n}=N-1$.

\subsection{The 6d origin of $\CS^{4d}_{(\varnothing, G)} \{SU(N), G \}/SU(N)$} 
The explicit F-theory quiver for \eref{6doriginS4d} is:
{\small
\be \label{6doriginS4dexp}
\begin{split}
G=SU(k): &\quad   \overset{\mathfrak{su_1}}2 \,\, \overset{\mathfrak{su_2}}2 \,\, \overset{\mathfrak{su_3}}2 \,\, \ldots \,\,  \overset{\mathfrak{su}_{k-1}}2 \,\, \underbrace{\underset{[N_f = 1]}{\overset{\mathfrak{su}_{k}}2}  \overset{\mathfrak{su}_{k}}2  \,\, \ldots \,\,\overset{\mathfrak{su}_{k}}2}_{N-k} \,\,  [SU(k)] \\
G=SO(2k): &\quad 2 \, \, \overset{\mathfrak{su_2}}2 \,\, \overset{\mathfrak{g_{2}}}3  \,\,  1 \,\,  \overset{\mathfrak{so_{9}}}4  \,\,   \overset{\mathfrak{usp_2}}1  \,\,    \cdots \,\,  \overset{\mathfrak{so}_{2k-1}}4  \,\, \underset{\left[N_f=\frac{1}{2} \right]}{\overset{\mathfrak{usp}_{2k-8}}1} \,\, \underbrace{\overset{\mathfrak{so}_{2k}}4 \bdash \overset{\mathfrak{so}_{2k}}4  \, \, \ldots \,\, \overset{\mathfrak{so}_{2k}}4}_{N-k} \bdash   [SO(2k)]\\
G= E_6: &\quad 2  \,\,   {\overset{\mathfrak{su_{2}}}2}  \,\,   {\overset{\mathfrak{g_2}}3}  \,\, 1 \,\, \overset{\mathfrak{f_{4}}}5 \,\, 1 \,\,  {\overset{\mathfrak{su_{3}}}3}  \,\,   1 \,\,\underbrace{\overset{\mathfrak{e_{6}}}6  \bdash  \overset{\mathfrak{e_{6}}}6 \bdash  \overset{\mathfrak{e_{6}}}6 \, \,  \ldots \, \, \overset{\mathfrak{e_{6}}}6}_{N-5} \bdash  [E_6] \\
G=E_7: &\quad 2 \,\, {\overset{\mathfrak{su_{2}}}2}\,\,  {\overset{\mathfrak{g_{2}}}3}    \,\, 1 \,\, {\overset{\mathfrak{f_{4}}}5}  \,\, 1\,\,  {\overset{\mathfrak{g_{2}}}3}  \,\, {\overset{\mathfrak{su_{2}}}2}  \,\, 1 \,\,  \underbrace{\overset{\mathfrak{e_{7}}}8  \bdash  \overset{\mathfrak{e_{7}}}8 \bdash  \overset{\mathfrak{e_{7}}}8 \, \,  \ldots \, \, \overset{\mathfrak{e_{7}}}8}_{N-5} \bdash [E_7] \\
G=E_8: &\quad 2 \,\, \overset{\mathfrak{su_{2}}}2 \,\, \overset{\mathfrak{g_{2}}}3  \,\, 1 \,\, \overset{\mathfrak{f_{4}}}5 \,\,  1 \,\, \overset{\mathfrak{g_{2}}}3 \,\, \overset{\mathfrak{su_{2}}}2 \,\, 2 \,\, 1\,\,  \overset{\mathfrak{e_{8}}}{11}  \,\,  1\,\, 2 \,\, \overset{\mathfrak{su_{2}}}2 \,\, \overset{\mathfrak{g_{2}}}3  \,\, 1 \,\, \overset{\mathfrak{f_{4}}}5 \,\,  1 \,\, \overset{\mathfrak{g_{2}}}3 \,\, \overset{\mathfrak{su_{2}}}2 \,\, 2 \,\, 1 \, \, \underbrace{ \overset{\mathfrak{e_{8}}}{12}  \bdash \overset{\mathfrak{e_{8}}}{12} \, \, \ldots \, \, \overset{\mathfrak{e_{8}}}{12}}_{N-6} \bdash  [E_8]~.
\end{split}
\ee}
The number under each of the braces indicates the number of copies of the minimal conformal matter theories (blue long dashes).  The CFT Higgs branch dimension of each theory can be computed using \eref{nform} as follows:
\be
\begin{split}
\dim^{\text{CFT}}_{\BH} \text{Higgs of \eref{6doriginS4dexp}} &= \mathfrak{n}'+\dim(G)+1 - \frac{1}{2}\left[ \dim(G)-\mathrm{rank}(G) \right]  \\
& \quad + (\dim(G) +1)(N - \mathfrak{n}'-1) - \dim(G) (N - \mathfrak{n}'-1) \\
&=  N + \frac{1}{2} \left[ \dim(G)+\mathrm{rank}(G) \right] ~.
\end{split}
\ee
Under the assumption that the $T^2$ compactification does not affect the Higgs branch moduli space, we claim that
\be \label{dimHiggsS4dG}
\dim_{\BH} \text{Higgs of $\frac{\CS^{4d}_{(\varnothing, G)} \{SU(N), G \}}{SU(N)}$} =  N + \frac{1}{2} \left[ \dim(G)+\mathrm{rank}(G) \right] ~.
\ee
In fact, it can be checked that this agrees with \eref{knowncompS4d} for $G=SU(k)$ and $G=SO(2k)$.  Using this dimension formula with \eref{T2compTbrane}, we find that
\be
\begin{split}
\dim^{\text{CFT}}_{\BH} \text{Higgs of \eref{T2compTbrane}}  &=  N + \frac{1}{2} \left[ \dim(G)+\mathrm{rank}(G) \right]  + \frac{3}{2} \left[ \dim(G)-\mathrm{rank}(G) \right] \nn \\
& \quad -d_{Y_{\rm L}}-d_{Y_{\rm R}} + \mathrm{rank}(G) - \dim(G) \\
&= N + \dim(G)-d_{Y_{\rm L}}-d_{Y_{\rm R}}~,
\end{split}
\ee
in agreement with \eref{nform} upon setting $\mathfrak{n}=N-1$.

It is worth noting that, as pointed out in \cite{Ohmori:2015pia}, for $G=SU(N)$, $Y_{\rm R} =[1^N]$ and $Y_{\rm L}$ a non-principal orbit, the $\diag(SU(N) \times SU(N))$ gauge group in \eref{T2compTbrane} is IR free. Even in this non-conformal case, our computation using the 4d description \eref{6doriginS4dexp} yields a result which is in agreement with the corresponding CFT Higgs branch dimension given by \eref{nform}.

\section{Conformal matter theories frozen to non-simply-laced groups on $T^2$} 
\label{sec:nonsimplylacedcm}

We will now consider partially frozen conformal matter theories --- more specifically, the HHN theories ${\cal T}^{\rm fr}_{G\to G_{\rm fr}}(N-1)$ discussed in section \ref{sec:M5}, which describe $N$ M5-branes probing a $\Gamma_G$ singularity frozen to $G_{\rm fr}$. When $G_{\rm fr}$ is a non-simply-laced Lie group, this is a possible definition of $(G_{\rm fr},G_{\rm fr})$ conformal matter. 

The resulting theories ${\cal T}^{\rm fr}_{G\to G_{\rm fr}}(0)$ were given in (\ref{minconfG2}). The longer chains ${\cal T}^{\rm fr}_{G\to G_{\rm fr}}(N)$ can once again be defined as $[G_{\rm fr}]\bdash [G_{\rm fr}]\bdash\ldots[G_{\rm fr}]$. We will shortly provide an explicit check that the CFT Higgs branch dimension of each of these theories is equal to $\dim(G_{\rm fr})+1$.  For $[SO(2k+1)] \bdash [SO(2k+1)]$ we do not manage to find a 6d F-theory quiver whose CFT Higgs branch dimension is equal to this value.\footnote{One may wish to consider the following quiver:
\be
{[SO(2k+1)]} \,\, \overset{\mathfrak{sp}_{k-3}}1  \,\, {\overset{\mathfrak{so}_{2k+3}}4} \, \, \,\, \underset{[N_f=\frac{1}{2}]}{\overset{\mathfrak{sp}_{k-2}}1}  \,\,  {\overset{\mathfrak{so}_{2k+4}}4}  \,\, \underset{[N_f=\frac{1}{2}]}{\overset{\mathfrak{sp}_{k-2}}1}   \,\, {\overset{\mathfrak{so}_{2k+3}}4} \,\, \overset{\mathfrak{sp}_{k-3}}1 \,\, [SO(2k+1)]~.
\ee
According to \eref{eq:dCFT}, this theory has the CFT Higgs branch dimension $2k^2+3k+6$, not $2k^2-k+1$.}

We also mentioned that these theories can be generalized by Higgsing the $G_{\rm fr}\times G_{\rm fr}$ flavor symmetry by two nilpotent elements, thus defining non-simply-laced T-brane theries. The cases $G_{\rm fr}=G_2$ and $F_4$ can be written explicitly using \cite[Sec.~4.2]{Heckman:2016ssk}.

We conjecture that the $T^2$ compactification of the CFT point of $\CT^{\rm fr}_{G\to G_{\rm fr}}(\{ \underline{Y}_L,\underline{Y}_R \}, \mathfrak{n})$ is described by
\be \label{T2compTbranenonsimp}
\frac{\CS^{4d}_{(\varnothing, \hat{G})}(SU(N),  \hat{G}) \times   \mathsf{S} \langle S^2 \rangle_{\hat{G}}\{[ 0_{\hat G}, \underline{Y}_L,\underline{Y}_R ] \}}{SU(N) \times \diag(\hat{G} \times \hat{G})}~,
\ee
where $\mathfrak{n} = N-1$; $\hat{G}$ denotes a simply laced group whose outer-automorphism action gives rise to the non-simply-laced group $G_{\rm fr}$;  $0_{\hat G}$ denotes the untwisted maximal puncture of $\hat{G}$;  $\underline{Y}_L$ and $\underline{Y}_R$ are twisted punctures of $\hat{G}$; and
\bi
\item  for $G_{\rm fr}=G_2$, we take $\hat{G}=SO(8)$; 
\item  for $G_{\rm fr}=F_4$, we take $\hat{G} = E_6$;
\item for $G_{\rm fr}=USp(2k)$, we take $\hat{G} =SO(2k+2)$;
\item for $G_{\rm fr}=SO(2k+1)$, we take $\hat{G} =SU(2k)$~.
\ei
Note that the Higgs branch dimension of $\CS^{4d}_{(\varnothing, \hat{G})}(SU(N),  \hat{G}) /SU(N)$ can be computed using \eref{dimHiggsS4dG}.   Below we compute the Higgs branch dimension of the 4d theory \eref{T2compTbranenonsimp} and compare with the CFT Higgs branch dimension of the 6d quivers in \eref{minconfG2}.  We find an agreement between the two results in each case.

\subsection{The case of $G=G_2$}  \label{sec:cmG2G2}

For ${\cal T}_{E_8\to G_2}(0)$ (minimal conformal matter of type $(G_2, G_2)$), the CFT Higgs branch dimension can be read off from (\ref{eq:dfcm}): it gives ${\rm dim}(G_2)+1=15$.

Upon compactifying the fixed point of this theory on $T^2$, we conjecture the resulting theory is the class $\CS$ theory of the twisted $SO(8)$ type associated with a sphere with two maximal $\BZ_3$-twisted punctures, denoted by $\underline{0}_{G_2}$ \cite{Chacaltana:2016shw}, and one (untwisted) subregular orbit $[5,3]$ of $SO(8)$:
\be \label{T2compG2G2}
\underline{0}_{G_2}~, \qquad \underline{0}_{G_2}~, \qquad [5,3]~.
\ee
For the trivial orbit $\underline{0}_{G_2}$, $n_H-n_V=112-107$; for the trivial orbit $[1^8]$ of $SO(8)$, $n_H-n_V=112-100$; and $11$ is the dimension of the $[5,3]$ orbit. Thus the Higgs branch dimension of this class $\CS$ theory is
\be
2(112-107-0)+(112-100-11)+4 = 15~,
\ee
in agreement with the six-dimensional result ${\rm dim}(G_2)+1=15$.

Now let us consider the Higgs branch dimension of the 4d system \eref{T2compTbranenonsimp}:
\be \label{dim4dcompG2}
\frac{\CS^{4d}_{(\varnothing, SO(8))}(SU(N),  SO(8)) \times   \mathsf{S} \langle S^2 \rangle_{SO(8)}\{ 0_{SO(8)}, \underline{Y}_L, \underline{Y}_R \}}{SU(N) \times SO(8)}~.
\ee
This is given by
\be
\begin{split}
&\dim_\BH~ \text{Higgs of \eref{dim4dcompG2}} \\
&= (N+16) + \Big[ (112-100-0)+ (112-107- \dim \, \underline{Y}_L) \\
& \qquad + (112-107- \dim \, \underline{Y}_R) +4 \Big] -28 \\
&= N+14 -\dim \, \underline{Y}_L -  \dim \, \underline{Y}_R\\
&= \mathfrak{n} + 15 -\dim \, \underline{Y}_L -  \dim \, \underline{Y}_R~,
\end{split}
\ee
which is in agreement with the CFT Higgs branch of $\CT^{6d}_{G_2}(\{Y_{\rm L}, Y_{\rm R} \}, \mathfrak{n})$ given by \eref{nform} with $G\to G_{\rm fr}$ (recall the difference between (\ref{eq:dCFTcm}) and (\ref{eq:dfcm})).

\subsection{The case of $G=F_4$}  \label{sec:cmF4F4}
For ${\cal T}_{E_8\to F_4}(0)$, minimal conformal matter of type $(F_4, F_4)$, the CFT Higgs branch dimension can be read off from (\ref{eq:dfcm}): it gives ${\rm dim}(F_4)+1=53$.

Upon compactifying the fixed point of this theory on $T^2$, we conjecture the resulting theory is the class $\CS$ theory of the twisted $E_6$ type associated with a sphere with two maximal twisted punctures, denoted by $\underline{0}_{F_4}$ \cite{Chacaltana:2015bna}, and one (untwisted) subregular orbit $E_6(a_1)$ of $E_6$:
\be \label{T2compF4F4}
\underline{0}_{F_4}~, \qquad \underline{0}_{F_4}~, \qquad E_6(a_1)~.
\ee
The Higgs branch dimension of this class $\CS$ theory can be computed as follows. 
$624-601$ is $n_H-n_V$ of the trivial orbit $\underline{0}_{F_4}$, $624-588$ is the $n_H-n_V$ of the trivial orbit $0$ of $E_6$, and $35$ is the dimension of the $E_6(a_1)$ orbit. Thus we get 
\be
2(624-601-0)+(624-588-35)+6 = 53 ~,
\ee
in agreement with the 6d result ${\rm dim}(F_4)+1=53$ above.

Now let us consider the Higgs branch dimension of the 4d system \eref{T2compTbranenonsimp}:
\be \label{dim4dcompF4}
\frac{\CS^{4d}_{(\varnothing, E_6)}(SU(N),  E_6) \times   \mathsf{S} \langle S^2 \rangle_{E_6}\{ 0_{E_6}, \underline{Y}_L, \underline{Y}_R \}}{SU(N) \times E_6}~,
\ee
This is given by
\be
\begin{split}
&\dim_\BH~ \text{Higgs of \eref{dim4dcompG2}} \\
&=  (N+42) + \Big[ (624 - 588-0)  \\
& \qquad + (624 - 601- \dim \, \underline{Y}_L) + (624-601- \dim \, \underline{Y}_R) +6 \Big] -78 \\
&= N+52 -\dim \, \underline{Y}_L -  \dim \, \underline{Y}_R\\
&= \mathfrak{n} + 53 -\dim \, \underline{Y}_L -  \dim \, \underline{Y}_R~,
\end{split}
\ee
which is in agreement with the CFT Higgs branch of $\CT^{6d}_{F_4}(\{Y_{\rm L}, Y_{\rm R} \}, \mathfrak{n})$ given by \eref{nform} with $G\to G_{\rm fr}$.

\subsection{The case of $G=USp(2k)$}
Once again, for ${\cal T}_{SO(2k+2)\to USp(2k)}(0)$, the CFT Higgs branch dimension can be read off from (\ref{eq:dfcm}): it gives
\be \label{dimminconfCN}
 \dim(USp(2k))+1 = 2k^2+k+1~.
\ee
Upon compactifying the fixed point of this theory on $T^2$, we conjecture the resulting theory is the class $\CS$ theory of the twisted $D_{k+1}$ type \cite{Chacaltana:2013oka} associated with a sphere with two maximal twisted punctures, denoted by $\underline{[1^{2k}]}$ and carrying flavour symmetry $USp(2k)$, and one (untwisted) subregular orbit $[2k-1,3]$ of $D_{k+1}$:
\be
\underline{[1^{2k}]}, \qquad \underline{[1^{2k}]}~, \qquad [2k-1,3]~.
\ee
The Higgs branch dimension of this class $\CS$ theory is
\be
\begin{split}
& 2\Big[ (k+1)^2-2 (k+1)+\frac{1}{2}-0 \Big]+[ (k+1)k \\ 
& \qquad - \{(k+1)^2 -(k+1)-1\} ]+ (k+1) \\
& = 2k^2+k+1 = \eref{dimminconfCN}~,
\end{split}
\ee
where $(k+1)^2-2 (k+1)+\frac{1}{2}$ is the value of $n_H-n_V$ for the puncture $\underline{[1^{2k}]}$; $(k+1)k$ is the value of $n_H-n_V$ for the puncture $[1^{2k+2}]$ of $D_{k+1}$; and  $\{(k+1)^2 -(k+1)-1\}$ is the dimension of the orbit $[2k-1,3]$ of $D_{k+1}$.

Now let us consider the Higgs branch dimension of the 4d system \eref{T2compTbranenonsimp}:
\be \label{dim4dcompCN}
\frac{\CS^{4d}_{(\varnothing, D_{k+1})}(SU(N),  D_{k+1}) \times   \mathsf{S} \langle S^2 \rangle_{D_{k+1}}\{ 0_{D_{k+1}}, \underline{Y}_L, \underline{Y}_R \}}{SU(N) \times SO(2k+2)}~,
\ee
This is given by
\be
\begin{split}
&\dim_\BH~ \text{Higgs of \eref{dim4dcompCN}} \\
&=  N+(k^2+1)  + \Big[ 2 \left\{ \frac{1}{2} - 2 (k + 1) + (k + 1)^2 \right \} +   (k + 1) k  \\
& \qquad  -  \dim \, \underline{Y}_L- \dim \, \underline{Y}_R + (k + 1) \Big] -\frac{1}{2}(2k+2)(2k+1) \\
&= N+\frac{1}{2}(2k)(2k+1) -\dim \, \underline{Y}_L -  \dim \, \underline{Y}_R\\
&= \mathfrak{n} + \dim(C_k) +1 -\dim \, \underline{Y}_L -  \dim \, \underline{Y}_R~,
\end{split}
\ee
which is in agreement with the CFT Higgs branch of $\CT^{6d}_{USp(2k)}(\{Y_{\rm L}, Y_{\rm R} \}, \mathfrak{n})$ given by \eref{nform} with $G\to G_{\rm fr}$.

\subsection{Remarks}

\paragraph{A puzzle about the Coulomb branch dimension} So far we have compared the CFT Higgs branch dimension of the 6d theory with the Higgs branch dimension of the 4d theory upon the $T^2$ compactification, and we have found a nice agreement between the two.  However, there is a mismatch when we compare (a) the number of the tensor multiplets + the total rank of the gauge groups in the minimal conformal matter of type $(G,G)$ with $G$ non-simply-laced with (b) the Coulomb branch dimension of the 4d theory upon the $T^2$ compactification.   

As an example, for the minimal conformal matter theory of type $(G_2, G_2)$, (a) the number of tensor multiplets + the total rank of the gauge groups is $11 + 1 + 8 + 1 + 2 + 4 = 27$, whereas (b) the Coulomb branch of 4d theory given by \eref{T2compG2G2} is $3$ complex dimensional (see $\# \, 18$ on Page 8 of \cite{Chacaltana:2016shw}).  There is mismatch of 24 complex degrees of freedom.

A possible explanation of this is that the result of compactification consists of an interacting SCFT and a collection of free vector multiplets.  In the above example, we conjecture that the aforementioned SCFT is the class $\CS$ theory \eref{T2compG2G2} and there are 24 free vector multiplets.  We hope to address this problem in future work.

\paragraph{The case of $G=SO(2k+1)$}
A frozen singularities of $\mathbb{C}^2/\Gamma_G$ in M-theory cannot possess $SO(2k+1)$ symmetry with $k\ge 4$ on it,
therefore there seems no natural way to define ``$(SO(2k+1), SO(2k+1))$ conformal matter".
Nevertheless, we can find that the Higgs branch dimensions of 4d systems \eref{T2compTbranenonsimp} with the formula \eref{nform} with $G=SO(2k+1)$ and nilpotent orbits $Y_L,Y_R$ of the group.
The $\widehat{G}$ in \eref{T2compTbranenonsimp} should be taken to be $SU(2k)$, so that the twisted full puncture $\underline{0}$ has $SO(2k+1)$ symmetry \cite{Chacaltana:2012ch}.
Then, the 4d system \eref{T2compTbranenonsimp} is
\be \label{dim4dcompBN}
\frac{\CS^{4d}_{(\varnothing, SU(2k))}(SU(N),  SU(2k)) \times   \mathsf{S} \langle S^2 \rangle_{SU(2k)}\{ [1^{2k}], \underline{Y}_L, \underline{Y}_R \}}{SU(N) \times SU(2k)}~,
\ee
and one can explicitly calculate the Higgs branch dimension which agrees with \eqref{nform}.

\subsection{Fractional M5-branes on $T^2$}
\label{sec:fracM5}

In this subsection, we will comment about the ``short'' theories discussed in section \ref{sub:partial}, namely the $(E_7, SO(7))$, $(E_8, G_2)$ and $(E_8, F_4)$ conformal matter theories, whose F-theory quivers we saw in (\ref{eq:short}). These theories describe $1/2$ M5-branes on $\BC^2/\Gamma_{E_7}$, $1/3$ M5-branes on $\BC^2/\Gamma_{E_8}$, and $1/2$ M5-branes on $\BC^2/\Gamma_{E_8}$, respectively. Although these are not really of the type ${\cal T}^{\rm fr}_{G\to G_{\rm fr}}(N)$, they are also frozen conformal matter theories in the sense of section \ref{sub:fm5}, and it turns out that we can check their dimensions as well.

The compactification on $T^2$ of $(E_7, SO(7))$ was studied in \cite{Ohmori:2015pua} and admits the class $\CS$ description as the $E_6$ type theory on a sphere with punctures $0, \, 2A_1, \, E_6(a_1)$.  The $T^2$ compactification of $(E_8, G_2)$ and $(E_8, F_4)$ were studied in \cite{Chacaltana:2017boe}.  In particular, $(E_8, G_2)$ admits the class $\CS$ description as the $E_7$ type theory on a sphere with punctures $0, \, 2A_2, \, E_7(a_1)$.  The $T^2$ compactification of $(E_8, F_4)$, on the other hand, is not found to have a direct class $\CS$ description but seem to appear in the class $\CS$ theory of the $E_8$ type associated with a sphere with punctures $0, \, D_4, \, E_8(a_1)$.  This class $\CS$ theory is in fact is a product of two SCFTs: one is the rank-1 Minahan-Nemeschansky $E_8$ SCFT and the other is an SCFT with an $E_8 \times F_4$ global symmetry.  The latter factor is conjectured to be $T^2$ compactification of $(E_8, F_4)$.  

The CFT Higgs branch dimension computed from \eref{eq:dCFT} agrees with the Higgs branch dimension of the 4d theory resulting from the $T^2$ compactification: they are $63$, $92$ and $136$ respectively.

\section{Completely frozen conformal matter on $T^3$} 
\label{sec:frozen}

In this section, we consider the quivers (\ref{frozenmin}) that arise from the completely frozen $D_4$, $E_6$, $E_7$ and $E_8$ singularities.

From (\ref{eq:dfcm}) with $G_{\rm fr}=$ trivial, we see that the CFT Higgs branch all of these quivers  has quaternionic dimension equal to 1, and hence isomorphic to $\BC^2/\Gamma$ for some finite subgroup $\Gamma$ of $SU(2)$; it is natural to conjecture that in fact $\Gamma= \Gamma_G$.  Upon compactifying on $T^2$ ({\it resp.} $T^3$), each of the resulting 4d ({\it resp.} 3d) theories has the complex ({\it resp.} quaternionic) Coulomb branch dimension equals to the number of tensor multiplets + the total rank of the gauge groups in the F-theory quiver. For all the theories in (\ref{frozenmin}), this turns out to be equal to
\begin{equation}\label{eq:coulombfrozen}
	h^\vee_G-1~,
\end{equation}
where $h^\vee_G$ is the dual Coxeter number of $G$. Note that (\ref{eq:coulombfrozen}) is also equal to the dimension of the reduced moduli space of one $G$ instanton on $\BC^2$.

Based on these properties, we conjecture that the $T^3$ compactification of each (\ref{frozenmin}) is a quiver with the shape of the affine Dynkin diagram of type $G$, with gauge groups being unitary groups of ranks equal to the Coxeter labels and overall $U(1)$ being removed \cite{Intriligator:1996ex, Cremonesi:2014xha}: see table \ref{tab:mirror}.

\begin{table}
\begin{center}
\begin{tabular}{|c|c|}
\hline
$G$ & Mirror theory \\
\hline
$SO(8)$ & $\node{}{U(1)}- \underset{\uer{}{U(1)}}{\node{\cver{}{U(1)}}{U(2)*}}-\node{}{U(1)}$ \\
\hline
$E_6$ &$\node{}{U(1)}-\node{}{U(2)}-\node{\overset{\cver{}{U(1)}}{\cver{}{U(2)}}}{U(3)*}-\node{}{U(2)}-\node{}{U(1)}$  \\
\hline
$E_7$ &$\node{}{U(1)}-\node{}{U(2)}-\node{}{U(3)}-\node{\cver{}{U(2)}}{U(4)*}-\node{}{U(3)}-\node{}{U(2)}-\node{}{U(1)} $\\
\hline
$E_8$  & $\node{}{U(1)}-\node{}{U(2)}- \cdots-\node{}{U(5)} -\node{\cver{}{U(3)}}{U(6)*}-\node{}{U(4)}-\node{}{U(2)}$ \\
\hline
\end{tabular}
\end{center}
\caption{Three dimensional mirror theory of the $T^3$ compactification of the theory on a single M5-brane on the completely frozen $\BC^2/\Gamma_G$ singularity.  Here $U(N)*$ denotes $U(N)/U(1) \cong SU(N)/\BZ_N$.}
\label{tab:mirror}
\end{table}
The $G=SO(8)$ case is confirmed by a result in \cite[Sec.~3.2]{DelZotto:2015rca}, according to which the $S^1$ compactification of this theory is an $SU(2)$ gauge group coupled to four copies of $\hat D_2(SU(2))$. Compactifying further to four dimensions, the latter become four copies of $D_2(SU(2))$, which is simply an $SU(2)$ fundamental hypermultiplet (see for example \cite[Eq.~(B.2.1),(10.5.2)]{tachikawa-review}). Upon further $S^1$ compactification down to three dimension, this should give the $SO(8)$ entry in table \ref{tab:mirror}.

The theories in table \ref{tab:mirror} are also the $3d$ mirror theories \cite{Intriligator:1996ex} of the $S^1$ reduction of the class $\CS$ theories of $SU(N)$ type associated with a sphere with 3 punctures \cite{Benini:2010uu}:
\be \label{classSinst}
\begin{split}
G=SO(8): &\qquad[1^2]~, \qquad [1^2]~, \qquad [1^2]~, \qquad [1^2]~;\\
G=E_6: &\qquad[1^3]~, \qquad [1^3]~, \qquad [1^3]~;\\
G=E_7: &\qquad[1^4]~, \qquad [2^2]~,\qquad [1^4]~; \\
G=E_8: &\qquad[1^5]~, \qquad [3^2]~, \qquad [2^3]~.
\end{split}
\ee

\paragraph{A duality chain.} For the compactification of the 6d theories in question on $T^3$, we can start from (A) an M5-brane probing $\BC^2/\Gamma_G$ and wrapping $T^3$.  Upon using the duality chain of \cite{Ohmori:2015pua}, we arrive at (B) an M2-brane probing $\BC^2/\Gamma_G \times T^3$.  A frozen  singularity in (A) is mapped to a commuting triple holonomy around $T^3$ far from the M2-brane in (B). In fact, the theories in table \ref{tab:mirror} were found in \cite{Porrati:1996xi} to describe the mirror of the worldvolume of an M2-brane probing $\BC^2/\Gamma_G \times \BC^2$ (whose direct description would be the ADHM field theory associated with the reduced moduli space of one $G$ instanton on $\BC^2$). It would be interesting to understand better the relationship between these two pictures.

\paragraph{A comment on the non-completely frozen cases.}  Let us briefly comment on $T^3$ compactification of $\CT_{\hat{G} \rightarrow G_{\mathrm{fr}}}(0)$ when $G_{\mathrm{fr}}$ is non-trivial. Recall that we have discussed $T^2$ compactification of such theories in section \ref{sec:nonsimplylacedcm}.  The result is a class $\CS$ theory of type $\hat{G}$ associated with a sphere with two twisted maximal puntures $\underline{0}_{G_{\mathrm{fr}}}$ and an untwisted minimal puncture $\hat{G}(a_1)$.  The 3d mirror of the $S^1$ compactification of this class $\CS$ theory is given by \cite{Benini:2010uu,Cremonesi:2014vla}
\be \label{3dcompo}
\frac{T_{\underline{0}_{G_{\mathrm{fr}}}}(G^\vee_{\mathrm{fr}}) \times T_{\underline{0}_{G_{\mathrm{fr}}}}(G^\vee_{\mathrm{fr}}) \times T_{\hat{G}(a_1)}(\hat{G})}{G^\vee_{\mathrm{fr}}/Z(G^\vee_{\mathrm{fr}})}~,
\ee
where $T_{\rho}(G)$, with $\rho$ an orbit of the Langlands dual group $G^\vee$ of $G$,  denotes a 3d $\CN=4$ theory living on the $1/2$-BPS domain wall in the 4d $\CN=4$ theory of gauge group $G$; this is described in \cite{Gaiotto:2008ak}.  The factor $G^\vee_{\mathrm{fr}}$ denominator denotes the gauging of the flavour symmetry $G^\vee_{\mathrm{fr}}$ coming from (1) two copies of $T_{\underline{0}_{G_{\mathrm{fr}}}}(G^\vee_{\mathrm{fr}})$, and (2) the $G^\vee_{\mathrm{fr}}$ subgroup of $\hat{G}$ flavour symmetry of $T_{\hat{G}(a_1)}(\hat{G})$.  The notation $Z(G)$ denotes the centre of the group $G$.   In particular, for $\hat{G} = SO(2k+2)$ and $G_{\mathrm{fr}} = USp(2k)$, so that $G^\vee_{\mathrm{fr}} = SO(2k+1)$, \eref{3dcompo} admits a Lagrangian description, which can be represented as a star-shaped quiver \cite{Benini:2010uu}:
\be \label{starshapefrozen}
\bnode{}{1} - \rnode{}{2}-\bnode{}{3} - \rnode{}{4} - \cdots - \bnode{}{2k-1} - \rnode{}{2k} - \bnode{\overset{\bluesqver{}{1}}{\rver{}{2}}}{2k+1} - \rnode{}{2k}- \bnode{}{2k-1} - \cdots  - \rnode{}{4}-\bnode{}{3}  - \rnode{}{2}- \bnode{}{1}
\ee
where the blue node with a label $m$ denotes the $O(m)$ gauge group and the red node with an even label $m$ denotes the $USp(m)$ gauge group. Each leg of the above quiver comes from two copies of the first of the following theories and one copy of the second of the following theories \cite{Gaiotto:2008ak, Cremonesi:2014kwa}:
\bea
T_{[1^{2k}]}(SO(2k+1)): &\quad \bnode{}{1} - \rnode{}{2}-\bnode{}{3} - \rnode{}{4} - \cdots - \bnode{}{2k-1} - \rnode{}{2k} - \sqbnode{}{2k+1}  \\
T_{[2k-1,3]}(SO(2k+2)): &\quad \rnode{}{2} - \sqbnode{}{2k+2} ~\text{or equivalently}~ \sqbnode{}{1}-\rnode{}{2} - \sqbnode{}{2k+1}~.
\eea
where the $SO(2k+1)$ flavour symmetries are commonly gauged and glued together.

Note that we have gauged the subgroup $SO(2k+1)$ of $SO(2k+2)$. One can check that the Coulomb branch dimension of \eref{starshapefrozen} agrees with \eref{dimminconfCN}:
\be
\begin{split}
\dim_\BH \, \text{Coulomb of \eref{starshapefrozen}} &= 2\left[\left( \sum_{j=0}^{k-1} j  \right) + \left( \sum_{j=1}^k j \right) \right]+k+1 \\
&= 2k^2+k+1= \eref{dimminconfCN}~.
\end{split}
\ee
The Higgs branch dimension of \eref{starshapefrozen} is
\be
\begin{split}
&\dim_\BH \, \text{Higgs of \eref{starshapefrozen}} \\
&=2 \left[ \sum_{j=1}^{2k} \frac{1}{2} j (j + 1)\right] + \frac{1}{2} (2) (2 k + 1) +\frac{1}{2}(2)(1)- 2 \left[ \sum_{j=0}^{k-1} \frac{1}{2} (2 j + 1) (2 j) \right] \\
& \quad  -2 \left[\sum_{j=1}^{k} \frac{1}{2} (2 j) (2 j + 1)\right]- \frac{1}{2} (2 k + 1) (2 k) - 3 \\
& = k-1~.
\end{split}
\ee
This agrees with the Coulomb branch dimension of the corresponding theory of class $\CS$; in particular, for $k=3$, the quaternionic Higgs branch dimension of \eref{starshapefrozen} is 2 which is equal to the sum of graded Coulomb branch dimensions of Fixture $\#66$ of \cite[p. 49]{Chacaltana:2013oka}.

\acknowledgments
We would like to express our thanks to Stefano Cremonesi, Amihay Hanany, Tom Rudelius, and Alberto Zaffaroni for useful discussions.  NM and AT are supported in part by the INFN.  NM also thanks the ERC Starting Grant 637844-HBQFTNCER and the organizers of the Pollica Summer Workshop.  AT is also supported in part by the ERC under Grant Agreement n.~307286 (XD-STRING). HS is partially supported by the Programs for Leading Graduate Schools, MEXT, Japan, via the Leading Graduate Course for Frontiers of Mathematical Sciences and Physics. HS is also supported by JSPS Research Fellowship for Young Scientists.

\bibliographystyle{ytphys}
\bibliography{ref,at}

\providecommand{\href}[2]{#2}\begingroup\raggedright\begin{thebibliography}{10}

\bibitem{intriligator-6d}
K.~A. Intriligator, ``{RG fixed points in six dimensions via branes at orbifold
  singularities},'' \href{http://dx.doi.org/10.1016/S0550-3213(97)00236-8}{{\em
  Nucl.Phys.} {\bfseries B496} (1997) 177--190},
\href{http://arxiv.org/abs/hep-th/9702038}{{\ttfamily arXiv:hep-th/9702038
  [hep-th]}}.

\bibitem{Intriligator:1997dh}
K.~A. Intriligator, ``{New string theories in six-dimensions via branes at
  orbifold singularities},'' {\em Adv. Theor. Math. Phys.} {\bfseries 1} (1998)
  271--282,
\href{http://arxiv.org/abs/hep-th/9708117}{{\ttfamily arXiv:hep-th/9708117}}.

\bibitem{DelZotto:2014hpa}
M.~Del~Zotto, J.~J. Heckman, A.~Tomasiello, and C.~Vafa, ``{6d Conformal
  Matter},'' \href{http://dx.doi.org/10.1007/JHEP02(2015)054}{{\em JHEP}
  {\bfseries 02} (2015) 054},
\href{http://arxiv.org/abs/1407.6359}{{\ttfamily arXiv:1407.6359 [hep-th]}}.

\bibitem{evans-johnson-shapere}
N.~J. Evans, C.~V. Johnson, and A.~D. Shapere, ``{Orientifolds, branes, and
  duality of 4D gauge theories},''
  \href{http://dx.doi.org/10.1016/S0550-3213(97)00384-2}{{\em Nucl. Phys.}
  {\bfseries B505} (1997) 251--271},
\href{http://arxiv.org/abs/hep-th/9703210}{{\ttfamily arXiv:hep-th/9703210}}.

\bibitem{Brunner:1997gf}
I.~Brunner and A.~Karch, ``{Branes at orbifolds versus Hanany Witten in
  six-dimensions},''
  \href{http://dx.doi.org/10.1088/1126-6708/1998/03/003}{{\em JHEP} {\bfseries
  03} (1998) 003},
\href{http://arxiv.org/abs/hep-th/9712143}{{\ttfamily arXiv:hep-th/9712143}}.

\bibitem{hanany-zaffaroni-6d}
A.~Hanany and A.~Zaffaroni, ``{Branes and six-dimensional supersymmetric
  theories},'' \href{http://dx.doi.org/10.1016/S0550-3213(98)00355-1}{{\em
  Nucl.Phys.} {\bfseries B529} (1998) 180--206},
\href{http://arxiv.org/abs/hep-th/9712145}{{\ttfamily arXiv:hep-th/9712145
  [hep-th]}}.

\bibitem{Gaiotto:2014lca}
D.~Gaiotto and A.~Tomasiello, ``{Holography for (1,0) theories in six
  dimensions},'' \href{http://dx.doi.org/10.1007/JHEP12(2014)003}{{\em JHEP}
  {\bfseries 12} (2014) 003},
\href{http://arxiv.org/abs/1404.0711}{{\ttfamily arXiv:1404.0711 [hep-th]}}.

\bibitem{Apruzzi:2013yva}
F.~Apruzzi, M.~Fazzi, D.~Rosa, and A.~Tomasiello, ``{All $AdS_7$ solutions of
  type II supergravity},''
  \href{http://dx.doi.org/10.1007/JHEP04(2014)064}{{\em JHEP} {\bfseries 04}
  (2014) 064},
\href{http://arxiv.org/abs/1309.2949}{{\ttfamily arXiv:1309.2949 [hep-th]}}.

\bibitem{10letter}
F.~Apruzzi, M.~Fazzi, A.~Passias, A.~Rota, and A.~Tomasiello,
  ``{Six-Dimensional Superconformal Theories and their Compactifications from
  Type IIA Supergravity},''
  \href{http://dx.doi.org/10.1103/PhysRevLett.115.061601}{{\em Phys. Rev.
  Lett.} {\bfseries 115} no.~6, (2015) 061601},
\href{http://arxiv.org/abs/1502.06616}{{\ttfamily arXiv:1502.06616 [hep-th]}}.

\bibitem{Cremonesi:2015bld}
S.~Cremonesi and A.~Tomasiello, ``{6d holographic anomaly match as a continuum
  limit},''
\href{http://arxiv.org/abs/1512.02225}{{\ttfamily arXiv:1512.02225 [hep-th]}}.

\bibitem{Cecotti:2010bp}
S.~Cecotti, C.~Cordova, J.~J. Heckman, and C.~Vafa, ``{T-Branes and
  Monodromy},'' \href{http://dx.doi.org/10.1007/JHEP07(2011)030}{{\em JHEP}
  {\bfseries 07} (2011) 030},
\href{http://arxiv.org/abs/1010.5780}{{\ttfamily arXiv:1010.5780 [hep-th]}}.

\bibitem{Heckman:2016ssk}
J.~J. Heckman, T.~Rudelius, and A.~Tomasiello, ``{6D RG Flows and Nilpotent
  Hierarchies},'' \href{http://dx.doi.org/10.1007/JHEP07(2016)082}{{\em JHEP}
  {\bfseries 07} (2016) 082},
\href{http://arxiv.org/abs/1601.04078}{{\ttfamily arXiv:1601.04078 [hep-th]}}.

\bibitem{Chacaltana:2012zy}
O.~Chacaltana, J.~Distler, and Y.~Tachikawa, ``{Nilpotent orbits and
  codimension-two defects of 6d N=(2,0) theories},''
  \href{http://dx.doi.org/10.1142/S0217751X1340006X}{{\em Int. J. Mod. Phys.}
  {\bfseries A28} (2013) 1340006},
\href{http://arxiv.org/abs/1203.2930}{{\ttfamily arXiv:1203.2930 [hep-th]}}.

\bibitem{Mekareeya:2016yal}
N.~Mekareeya, T.~Rudelius, and A.~Tomasiello, ``{T-branes, Anomalies and Moduli
  Spaces in 6D SCFTs},''
\href{http://arxiv.org/abs/1612.06399}{{\ttfamily arXiv:1612.06399 [hep-th]}}.

\bibitem{Ohmori:2015pua}
K.~Ohmori, H.~Shimizu, Y.~Tachikawa, and K.~Yonekura, ``{6d $\mathcal{N}=(1,0)$
  theories on $T^2$ and class S theories: Part I},''
  \href{http://dx.doi.org/10.1007/JHEP07(2015)014}{{\em JHEP} {\bfseries 07}
  (2015) 014},
\href{http://arxiv.org/abs/1503.06217}{{\ttfamily arXiv:1503.06217 [hep-th]}}.

\bibitem{witten-without}
E.~Witten, ``{Toroidal compactification without vector structure},''
  \href{http://dx.doi.org/10.1088/1126-6708/1998/02/006}{{\em JHEP} {\bfseries
  02} (1998) 006},
\href{http://arxiv.org/abs/hep-th/9712028}{{\ttfamily arXiv:hep-th/9712028
  [hep-th]}}.

\bibitem{deBoer:2001px}
J.~de~Boer, R.~Dijkgraaf, K.~Hori, A.~Keurentjes, J.~Morgan, D.~R. Morrison,
  and S.~Sethi, ``{Triples, Fluxes, and Strings},'' {\em Adv. Theor. Math.
  Phys.} {\bfseries 4} (2002) 995--1186,
\href{http://arxiv.org/abs/hep-th/0103170}{{\ttfamily arXiv:hep-th/0103170}}.

\bibitem{tachikawa-frozen}
Y.~Tachikawa, ``{Frozen},''
\href{http://arxiv.org/abs/1508.06679}{{\ttfamily arXiv:1508.06679 [hep-th]}}.

\bibitem{Heckman:2013pva}
J.~J. Heckman, D.~R. Morrison, and C.~Vafa, ``{On the Classification of 6D
  SCFTs and Generalized ADE Orbifolds},''
  \href{http://dx.doi.org/10.1007/JHEP06(2015)017,
  10.1007/JHEP05(2014)028}{{\em JHEP} {\bfseries 05} (2014) 028},
  \href{http://arxiv.org/abs/1312.5746}{{\ttfamily arXiv:1312.5746 [hep-th]}}.
[Erratum: {\textit{JHEP}} {$\mathbf{06}$} (2015) 017].

\bibitem{Ohmori:2015pia}
K.~Ohmori, H.~Shimizu, Y.~Tachikawa, and K.~Yonekura, ``{6d
  $\mathcal{N}{=}(1,0)$ theories on $S^1/T^2$ and class S theories: part II},''
\href{http://arxiv.org/abs/1508.00915}{{\ttfamily arXiv:1508.00915 [hep-th]}}.

\bibitem{Ohmorithesis2015}
K.~Ohmori, {\em {Six-Dimensional Superconformal Field Theories and Their Torus
  Compactifications}}.
\newblock PhD thesis, {University of Tokyo}, 2016.
\newblock
  \url{http://www2.yukawa.kyoto-u.ac.jp/~soken.editorial/shuron/ohmori_k16.pdf}.

\bibitem{Heckman:2015bfa}
J.~J. Heckman, D.~R. Morrison, T.~Rudelius, and C.~Vafa, ``{Atomic
  Classification of 6D SCFTs},''
  \href{http://dx.doi.org/10.1002/prop.201500024}{{\em Fortsch. Phys.}
  {\bfseries 63} (2015) 468--530},
\href{http://arxiv.org/abs/1502.05405}{{\ttfamily arXiv:1502.05405 [hep-th]}}.

\bibitem{Bena:2016oqr}
I.~Bena, J.~{Bl\aa b\"ack}, R.~Minasian, and R.~Savelli, ``{There and back
  again: A T-brane's tale},''
  \href{http://dx.doi.org/10.1007/JHEP11(2016)179}{{\em JHEP} {\bfseries 11}
  (2016) 179},
\href{http://arxiv.org/abs/1608.01221}{{\ttfamily arXiv:1608.01221 [hep-th]}}.

\bibitem{Ohmori:2014kda}
K.~Ohmori, H.~Shimizu, Y.~Tachikawa, and K.~Yonekura, ``{Anomaly polynomial of
  general 6d SCFTs},'' \href{http://dx.doi.org/10.1093/ptep/ptu140}{{\em PTEP}
  {\bfseries 2014} no.~10, (2014) 103B07},
\href{http://arxiv.org/abs/1408.5572}{{\ttfamily arXiv:1408.5572 [hep-th]}}.

\bibitem{Hanany:2010qu}
A.~Hanany and N.~Mekareeya, ``{Tri-vertices and SU(2)'s},''
  \href{http://dx.doi.org/10.1007/JHEP02(2011)069}{{\em JHEP} {\bfseries 02}
  (2011) 069},
\href{http://arxiv.org/abs/1012.2119}{{\ttfamily arXiv:1012.2119 [hep-th]}}.

\bibitem{Chacaltana:2012ch}
O.~Chacaltana, J.~Distler, and Y.~Tachikawa, ``{Gaiotto Duality for the Twisted
  $A_{2N-1}$ Series},''
\href{http://arxiv.org/abs/1212.3952}{{\ttfamily arXiv:1212.3952 [hep-th]}}.

\bibitem{Benini:2010uu}
F.~Benini, Y.~Tachikawa, and D.~Xie, ``{Mirrors of 3d Sicilian theories},''
  \href{http://dx.doi.org/10.1007/JHEP09(2010)063}{{\em JHEP} {\bfseries 09}
  (2010) 063},
\href{http://arxiv.org/abs/1007.0992}{{\ttfamily arXiv:1007.0992 [hep-th]}}.

\bibitem{Cremonesi:2014vla}
S.~Cremonesi, A.~Hanany, N.~Mekareeya, and A.~Zaffaroni, ``{Coulomb branch
  Hilbert series and Three Dimensional Sicilian Theories},''
  \href{http://dx.doi.org/10.1007/JHEP09(2014)185}{{\em JHEP} {\bfseries 09}
  (2014) 185},
\href{http://arxiv.org/abs/1403.2384}{{\ttfamily arXiv:1403.2384 [hep-th]}}.

\bibitem{Chacaltana:2016shw}
O.~Chacaltana, J.~Distler, and A.~Trimm, ``{Tinkertoys for the Z3-twisted D4
  Theory},''
\href{http://arxiv.org/abs/1601.02077}{{\ttfamily arXiv:1601.02077 [hep-th]}}.

\bibitem{Chacaltana:2015bna}
O.~Chacaltana, J.~Distler, and A.~Trimm, ``{Tinkertoys for the Twisted $E_6$
  Theory},'' \href{http://dx.doi.org/10.1007/JHEP04(2015)173}{{\em JHEP}
  {\bfseries 04} (2015) 173},
\href{http://arxiv.org/abs/1501.00357}{{\ttfamily arXiv:1501.00357 [hep-th]}}.

\bibitem{Chacaltana:2013oka}
O.~Chacaltana, J.~Distler, and A.~Trimm, ``{Tinkertoys for the Twisted
  D-Series},''
\href{http://arxiv.org/abs/1309.2299}{{\ttfamily arXiv:1309.2299 [hep-th]}}.

\bibitem{Chacaltana:2017boe}
O.~Chacaltana, J.~Distler, A.~Trimm, and Y.~Zhu, ``{Tinkertoys for the E7
  Theory},''
\href{http://arxiv.org/abs/1704.07890}{{\ttfamily arXiv:1704.07890 [hep-th]}}.

\bibitem{Intriligator:1996ex}
K.~A. Intriligator and N.~Seiberg, ``{Mirror symmetry in three-dimensional
  gauge theories},'' \href{http://dx.doi.org/10.1016/0370-2693(96)01088-X}{{\em
  Phys. Lett.} {\bfseries B387} (1996) 513--519},
\href{http://arxiv.org/abs/hep-th/9607207}{{\ttfamily arXiv:hep-th/9607207
  [hep-th]}}.

\bibitem{Cremonesi:2014xha}
S.~Cremonesi, G.~Ferlito, A.~Hanany, and N.~Mekareeya, ``{Coulomb Branch and
  The Moduli Space of Instantons},''
  \href{http://dx.doi.org/10.1007/JHEP12(2014)103}{{\em JHEP} {\bfseries 12}
  (2014) 103},
\href{http://arxiv.org/abs/1408.6835}{{\ttfamily arXiv:1408.6835 [hep-th]}}.

\bibitem{DelZotto:2015rca}
M.~Del~Zotto, C.~Vafa, and D.~Xie, ``{Geometric Engineering, Mirror Symmetry
  and 6d (1,0) -> 4d, N=2},''
\href{http://arxiv.org/abs/1504.08348}{{\ttfamily arXiv:1504.08348 [hep-th]}}.

\bibitem{tachikawa-review}
Y.~Tachikawa, \href{http://dx.doi.org/10.1007/978-3-319-08822-8}{{\em {${\cal
  N}=2$ supersymmetric dynamics for pedestrians}}}, vol.~890.
\newblock 2014.
\newblock
\href{http://arxiv.org/abs/1312.2684}{{\ttfamily arXiv:1312.2684 [hep-th]}}.
\newblock

\bibitem{Porrati:1996xi}
M.~Porrati and A.~Zaffaroni, ``{M theory origin of mirror symmetry in
  three-dimensional gauge theories},''
  \href{http://dx.doi.org/10.1016/S0550-3213(97)00061-8}{{\em Nucl.Phys.}
  {\bfseries B490} (1997) 107--120},
\href{http://arxiv.org/abs/hep-th/9611201}{{\ttfamily arXiv:hep-th/9611201
  [hep-th]}}.

\bibitem{Gaiotto:2008ak}
D.~Gaiotto and E.~Witten, ``{S-Duality of Boundary Conditions In N=4 Super
  Yang-Mills Theory},''
  \href{http://dx.doi.org/10.4310/ATMP.2009.v13.n3.a5}{{\em
  Adv.Theor.Math.Phys.} {\bfseries 13} (2009) 721},
\href{http://arxiv.org/abs/0807.3720}{{\ttfamily arXiv:0807.3720 [hep-th]}}.

\bibitem{Cremonesi:2014kwa}
S.~Cremonesi, A.~Hanany, N.~Mekareeya, and A.~Zaffaroni, ``{Coulomb branch
  Hilbert series and Hall-Littlewood polynomials},''
  \href{http://dx.doi.org/10.1007/JHEP09(2014)178}{{\em JHEP} {\bfseries 09}
  (2014) 178},
\href{http://arxiv.org/abs/1403.0585}{{\ttfamily arXiv:1403.0585 [hep-th]}}.

\end{thebibliography}\endgroup

\end{document}